\RequirePackage[british]{babel}
\documentclass[reqno,a4paper,12pt]{amsart}
\usepackage[a4paper,hmargin=2cm,tmargin=3cm,bmargin=3cm]{geometry}
\usepackage[utf8x]{inputenc}
\usepackage{amsmath,amssymb,amstext,amsthm,amscd,mathrsfs,eucal}
\usepackage{graphicx,color}
\usepackage{array}
\usepackage{cite}
\usepackage{hyperref}
\hypersetup{%
  pdftitle   = {Deformations of 3-algebras},
  pdfkeywords = {Leibniz algebra, Lie algebra, n-Lie algebra,
    n-Leibniz algebra, deformations, cohomology,
    representations},
  pdfauthor  = {José Figueroa-O'Farrill},
  pdfcreator = {\LaTeX\ with package \flqq hyperref\frqq}
}
\PrerenderUnicode{éÉ}
\let\into\hookrightarrow

\newcommand{\half}{\tfrac12}
\newcommand{\fg}{\mathfrak{g}}

\newcommand{\fM}{\mathfrak{M}}

\newcommand{\fgl}{\mathfrak{gl}}

\newcommand{\fs}{\mathfrak{s}}

\newcommand{\fso}{\mathfrak{so}}

\newcommand{\SO}{\mathrm{SO}}

\newcommand{\GL}{\mathrm{GL}}

\newcommand{\RR}{\mathbb{R}}

\newcommand{\ZZ}{\mathbb{Z}}
\newcommand{\eD}{\mathscr{D}}

\newcommand{\1}{\boldsymbol{1}}
\DeclareMathOperator{\Aut}{Aut}

\DeclareMathOperator{\Hom}{Hom}
\DeclareMathOperator{\End}{End}

\DeclareMathOperator{\rad}{rad}

\theoremstyle{plain}
\newtheorem{lemma}{Lemma}
\newtheorem{proposition}[lemma]{Proposition}
\newtheorem{theorem}[lemma]{Theorem}

\theoremstyle{definition}

\newtheorem{example}[lemma]{Example}
\newcommand{\MUNCH}[1]{\relax}

\allowdisplaybreaks
\begin{document}
\title{Deformations of 3-algebras}
\author[José Miguel Figueroa-O'Farrill]{José Miguel Figueroa-O'Farrill}
\address{Maxwell Institute and School of Mathematics, University of
  Edinburgh, Scotland, UK}
\email{J.M.Figueroa@ed.ac.uk}
\begin{abstract}
  We phrase 
  deformations of $n$-Leibniz algebras in terms of the cohomology
  theory of the associated Leibniz algebra.  We do the same for
  $n$-Lie algebras and for the metric versions of $n$-Leibniz and
  $n$-Lie algebras.  We place particular emphasis on the case of $n=3$
  and explore the deformations of 3-algebras of relevance to
  three-dimensional superconformal Chern--Simons theories with
  matter.
\end{abstract}
\maketitle
\tableofcontents

\section{Introduction}
\label{sec:introduction}

It is often said that one can learn a lot about a mathematical object
by studying how it behaves under deformations.  In this paper we will
study the deformations of certain algebraic structures which have
appeared recently in the Bagger--Lambert--Gustavsson theory of
multiple M2-branes \cite{BL1,GustavssonAlgM2,BL2}.  More generally,
they underlie certain three-dimensional superconformal Chern--Simons
theories coupled to matter, where the matter fields take values in a
metric 3-Leibniz algebra (see below for a precise definition).  Not
every 3-Leibniz algebra is associated to one such theory, but those
which are can be constructed from a metric Lie algebra and a faithful
unitary representation \cite{Lie3Algs} via a construction originally
due to Faulkner \cite{FaulknerIdeals}.  A special class of such
algebras are the metric 3-Lie algebras, which go back to the work of
Filippov

Let $V$ be a real $n$-Lie algebra \cite{Filippov}; that is, a real
vector space and an alternating multilinear map $V^n \to V$ denoted by
\begin{equation}
  (x_1,\dots,x_n) \mapsto [x_1,\dots,x_n]
\end{equation}
satisfying a \emph{fundamental identity} which says that the
endomorphisms of $V$ defined by $y \mapsto [x_1,\dots,x_{n-1},y]$ are
derivations for all $x_i \in V$.  This suggests defining a linear map
$D: \Lambda^{n-1} V \to \End V$ by
\begin{equation}
  D(x_1 \wedge \dots \wedge x_{n-1})\cdot y = [x_1,\dots,x_{n-1},y]
\end{equation}
and extending it linearly to all of $\Lambda^{n-1}V$.  In terms of
this map, the fundamental identity can be written as a simple relation
in $\End V$.
\begin{lemma}
  The fundamental identity is equivalent to 
  \begin{equation}
    \label{eq:FI}
    [D(X), D(Y)] = D(D(X)\cdot Y)~,
  \end{equation}
  for all $X,Y \in \Lambda^{n-1}V$, where the $\cdot$ in the
  right-hand side is the natural action of $\End V$ on $\Lambda^{n-1}V$
  and the bracket on the left-hand side is the commutator in $\End V$.
\end{lemma}

\begin{proof}
  Indeed, let $X = x_1\wedge\dots\wedge x_{n-1}$ and $Y=
  y_1\wedge\dots\wedge y_{n-1}$ be monomials.  Then
  \begin{align*}
    D(X) \cdot Y &= D(x_1\wedge\dots\wedge x_{n-1}) \cdot
    (y_1\wedge\dots\wedge y_{n-1})\\
    &= \sum_{i=1}^{n-1} y_1 \wedge \dots \wedge [x_1,\dots,x_{n-1},y_i]
    \wedge \dots y_{n-1}~,
  \end{align*}
  whence applying the fundamental identity to $y_n \in V$, we find
  \begin{align*}
    [D(X), D(Y)] \cdot y_n &= [x_1,\dots,x_{n-1},[y_1,\dots,y_{n-1},y_n]] -
    [y_1,\dots,y_{n-1},[x_1,\dots,x_{n-1},y_n]] \\
    \intertext{whereas}
    D(D(X)\cdot Y) \cdot y_n &= \sum_{i=1}^{n-1} D(y_1 \wedge \dots \wedge [x_1,\dots,x_{n-1},y_i]
    \wedge \dots y_{n-1}) \cdot y_n \\
    &= \sum_{i=1}^{n-1} [y_1, \dots, [x_1,\dots,x_{n-1},y_i], \dots,
    y_{n-1}, y_n]~.
  \end{align*}
  Into the fundamental identity and rearranging terms we obtain
  \begin{equation}
    [x_1,\dots,x_{n-1},[y_1,\dots,y_{n-1},y_n]]  = \sum_{i=1}^n [y_1,
    \dots, [x_1,\dots,x_{n-1},y_i], \dots, y_{n-1}, y_n]~,
  \end{equation}
  which says precisely that $D(x_1\wedge \dots\wedge x_{n-1})$ is a
  derivation over the bracket.
\end{proof}

The fundamental identity in the form \eqref{eq:FI} says that the image
of $D$ is a Lie subalgebra of $\fgl(V)$.  At first sight, this is
unexpected, because $D$ is only a linear map and hence there is no
right to expect that its image should be a Lie subalgebra.  A natural
explanation for this fact would be that $D$ is a Lie algebra
morphism, but this would require $\Lambda^{n-1} V$ to possess the
structure of a Lie algebra.  Taking this seriously and glancing at
equation \eqref{eq:FI}, we would be tempted to define the bracket in
$\Lambda^{n-1}V$ by
\begin{equation}
  \label{eq:Leib1}
  [X,Y] := D(X) \cdot Y~,
\end{equation}
in such a way that the fundamental identity \eqref{eq:FI} becomes
\begin{equation}
  \label{eq:FI-L-morph}
  [D(X),D(Y)] = D([X,Y])~.
\end{equation}

Although the bracket defined by \eqref{eq:Leib1} is not skewsymmetric,
it does however obey a version of the Jacobi identity:
\begin{equation}
  \label{eq:LI}
  [X, [Y, Z]] = [[X,Y],Z] + [Y,[X,Z]]~,
\end{equation}
for all $X,Y,Z \in \Lambda^{n-1} V$.  (We will prove this later in a
slightly more general context.)  It turns out that this makes
$\Lambda^{n-1}V$ into a \emph{(left) Leibniz algebra} (see below for a
precise definition) and then the fundamental identity in the form
\eqref{eq:FI-L-morph} makes $D$ into a (left) Leibniz algebra
morphism, whence explaining satisfactorily the fact that its image is
a Leibniz (and hence Lie) subalgebra of $\fgl(V)$.  (This observation
seems to have been made by Daletskii in the first instance.)

It turns out that this is the correct framework in which to discuss
the cohomology of $n$-Lie algebras.  In particular we will show that
deformations of an $n$-Lie algebra are governed by certain cohomology
groups of the associated Leibniz algebra $\Lambda^{n-1}V$.  Leibniz
algebras were introduced by Loday in \cite{MR1217970} and the
cohomology was discussed by Cuvier in \cite{MR1133486,MR1258404} for
the special case of symmetric representations (see below for a precise
definition) and by Loday and Pirashvili in \cite{MR1213376} in
general.  The question of deformations of $n$-Lie algebras (motivated
by the quantisation problem for Nambu mechanics \cite{Nambu:1973qe})
was tackled successfully by Gautheron in \cite{MR1392151}, after an
initial attempt by Takhtajan in \cite{MR1290830}.  Daletskii and
Takhtajan in \cite{MR1437747} rewrote Gautheron's result in terms of
Leibniz cohomology.  One purpose of this paper is to obtain a clearer
statement of which cohomology groups control the deformations of an
$n$-Leibniz algebra, of which an $n$-Lie algebra is a special case.

This paper is organised as follows.  The first two sections are mostly
a review of known results, collected here under one roof.  In
Section~\ref{sec:algebras} we recall the definition of a Leibniz
algebra and its associated Lie algebra.  We introduce the notion of a
representation and in particular the special cases of symmetric and
antisymmetric representations.  We recall the definition of the
universal enveloping algebra of a Leibniz algebra and of the
differential complex computing the cohomology $HL^*(L;M)$ of a (left)
Leibniz algebra $L$ with values in a representation $M$.  In
Section~\ref{sec:LnA} we introduce the notion of an $n$-Leibniz
algebra and describe a functor which assigns a (metric) Leibniz
algebra canonically to each such (metric) $n$-Leibniz algebra.
In Section~\ref{sec:deformations} we show that infinitesimal
deformations of $V$ (in the sense of Gerstenhaber \cite{Gerstenhaber})
are classified by $HL^1(L;\End V)$, where $\End V$ is a
\emph{nonsymmetric} representation of $L$, whereas the obstructions to
integrating an infinitesimal deformation live in $HL^2(L;\End V)$.  In
order to arrive at this result we endow $CL^\bullet(L;\End V)$ with a
graded Lie algebra structure as in the Nijenhuis--Richardson theory of
Lie algebra deformations \cite{NijenhuisRichardson}.  We also discuss
the deformation theory of $n$-Lie algebras and of metric $n$-Leibniz
and $n$-Lie algebras.  In Section~\ref{sec:n=3} we specialise to the
3-Leibniz algebras of interest in the context of three-dimensional
superconformal Chern--Simons theories.  As shown in \cite{Lie3Algs},
these are in one-to-one correspondence with pairs consisting of a
metric Lie algebra and a faithful orthogonal representation, via a
construction originally due to Faulkner \cite{FaulknerIdeals}.  We
discuss the relevant Leibniz algebra in the general Faulkner
construction and and set up the deformation theory in the special case
of orthogonal representations.  We show that the unique simple
euclidean 3-Lie algebra is rigid as a 3-Lie algebra, but admits a
one-parameter family of nontrivial deformations as a 3-Leibniz
algebra, which we interpret in terms of deformations of its Faulkner
data.  Finally we discuss the deformation problem for general Faulkner
data and show that if the metric Lie algebra in the Faulkner
construction is semisimple, then the deformations all correspond to
rescaling the Killing forms in each of its simple ideals.

\section{Leibniz algebras, their representations and their cohomology}
\label{sec:algebras}

In this section we recall some definitions in the theory of Leibniz
algebras and introduce the notion of representation, universal
enveloping algebra and cohomology with values in a representation.
The treatment follows \cite{MR1213376}.

\subsection{Basic definitions}
\label{sec:basic-definitions}

A \textbf{(left) Leibniz algebra} is a vector space $L$ together
with a bilinear map $L^2 \to L$, $(X,Y)\mapsto [X,Y]$ satisfying the
\textbf{(left) Leibniz identity}:
\begin{equation}
  \label{eq:LLI}
  [X,[Y,Z]] = [[X,Y],Z] + [Y,[X,Z]]~, \qquad\text{for all $X,Y,Z \in L$.}
\end{equation}
If the bracket were skewsymmetric, so that $[X,Y] = - [Y,X]$, then the
Leibniz identity would become the Jacobi identity, making $L$ into a
Lie algebra.  In a sense, Leibniz algebras are noncommutative versions
of Lie algebras.  Because of this noncommutativity one has to
distinguish between left and right Leibniz algebras.  Some of the
relevant literature (e.g., \cite{MR1258404,MR1213376}) considers right
Leibniz algebras, whereas in the present context the natural notion is
that of a left Leibniz algebra defined above.  Care must be exercised
in translating from one to the other.  All our Leibniz algebras will
be left Leibniz algebras unless otherwise stated.

Let $K\subset L$ denote the subspace of $L$ spanned by
$\left\{[X,X]\middle | X \in L\right\}$.

\begin{lemma}
  If $Z \in K$, then $[Z,X]=0$ and $[X,Z] \in K$, for all $X \in L$.
\end{lemma}

\begin{proof}
  This follows from the Leibniz identity \eqref{eq:LLI}.  Indeed,
  \begin{equation}
    [[X,X],Y] = [X,[X,Y]] - [X,[X,Y]] = 0~,
  \end{equation}
  and
  \begin{equation}
    [Y,[X,X]] = [[Y,X],X] + [X,[Y,X]]~,
  \end{equation}
  but we notice that any element of the form $[X,Y] + [Y,X]$ belongs to
  $K$, since
  \begin{equation}
    [X,Y] + [Y,X] = \half[X+Y,X+Y] - \half [X-Y,X-Y]~.
  \end{equation}
\end{proof}

In other words, $K$ is an ideal of $L$ and the quotient $\fg_L := L/K$
is therefore a Leibniz algebra whose bracket is skewsymmetric by
construction.  This makes $\fg_L$ into a Lie algebra known as the Lie
algebra associated to the Leibniz algebra $L$.  If $X \in L$, we will
let $\overline X \in \fg_L$ denote the corresponding element.  If
$\fg$ is a Lie algebra, thought of as a Leibniz algebra, then any
Leibniz algebra morphism $L \to \fg$ factors through a unique Lie
algebra morphism $\fg_L \to \fg$.  In this sense, $\fg_L$ is universal
for Leibniz morphisms to Lie algebras.

\subsection{Representations and the universal enveloping algebra}
\label{sec:representations}

In this section we translate results from \cite{MR1213376} from right
to left Leibniz algebras.

An \textbf{abelian extension} of a Leibniz algebra $L$ is an exact
sequence
\begin{equation}
  \label{eq:abext}
  \begin{CD}
    0 @>>> M @>>> L' @>>> L @>>> 0
  \end{CD}
\end{equation}
of Leibniz algebras where $[M,M]=0$.  This means that $M$ admits two
actions of $L$: a left action $L \times M \to M$ denoted $[X,m]$ and a
right action $M \times L \to M$, denoted $[m,X]$ for $X\in L$ and
$m\in M$.  These actions are given by the bracket of $L'$ after
choosing \emph{any} vector space section $L \to L'$ and as a result
satisfy three compatibility conditions coming from the Leibniz
identity in $L'$ applied to two elements of $L$ and one of $M$ in
different positions, namely:
\begin{align}
  [[X,Y],m] &= [X,[Y,m]] - [Y,[X,m]]   \label{eq:LLM}\\
  [[X,m],Y] &= [X,[m,Y]] - [m,[X,Y]]   \label{eq:LML}\\
  [[m,X],Y] &= [m,[X,Y]] - [X,[m,Y]]~,   \label{eq:MLL}
\end{align}
for all $X,Y\in L$ and $m \in M$.  Notice that adding the last two
equations, we have
\begin{equation}
  \label{eq:Z}
  [[X,m],Y] + [[m,X],Y] = 0~.
\end{equation}
We say that a vector space $M$ is a \textbf{representation} of the
Leibniz algebra $L$ if $M$ admits two actions of $L$, on the left and
on the right, obeying equations \eqref{eq:LLM}, \eqref{eq:LML} and
\eqref{eq:MLL}.  A vector subspace $N\subset M$ is a
\textbf{subrepresentation} if it is closed under both the left and
right actions of $L$.  A representation $M$ is said to be
\textbf{symmetric} if $[X,m] + [m,X] = 0$ for all $X \in L$ and $m \in
M$.  In this case, equations \eqref{eq:LML} and \eqref{eq:MLL} are
equivalent to equation \eqref{eq:LLM}, which we may take as the
defining condition for a symmetric representation.  Similarly, a
representation is said to be \textbf{antisymmetric} if $[m,X] = 0$.
In this case, the only nontrivial condition is again equation
\eqref{eq:LLM}.

As for a Lie algebra, a Leibniz algebra has a universal enveloping
algebra: an associative algebra such that its (left) modules are in
one-to-one correspondence with the representations of the Leibniz
algebra.  Because representations of a Leibniz algebra consists of
left and right actions, the universal enveloping algebra of a Leibniz
algebra $L$ is a quotient of the tensor algebra of $L\oplus L$.  For
$X\in L$, we let $r_X = (0,X)$ and $\ell_X = (X,0)$ denote the
corresponding elements in $L \oplus L$.  The \textbf{universal
  enveloping algebra} $UL(L)$ of $L$ is the quotient of the tensor
algebra $T(L\oplus L)$ by the two-sided ideal $I$ generated by the
following elements
\begin{equation}
  \label{eq:relations}
  \ell_{[X,Y]}- \ell_X\otimes\ell_Y + \ell_Y \otimes \ell_X
  \qquad
  r_{[X,Y]} - \ell_X \otimes r_Y - r_Y \otimes \ell_X
  \qquad
  r_Y \otimes (\ell_X + r_X)~,
\end{equation}
for all $X,Y\in L$.

\begin{proposition}
  There is a categorical equivalence between representations of $L$
  and left modules of $UL(L)$.
\end{proposition}

\begin{proof}
  If $M$ is a representation of $L$ we define
  \begin{equation}
    \ell_X m := [X,m]  \qquad\text{and}\qquad r_X m := [m,X]
  \end{equation}
  for all $X\in L$ and $m\in M$.  We extend it to all of $T(L \oplus L)$
  by linearity and composition.  The conditions \eqref{eq:LLM},
  \eqref{eq:LML} and \eqref{eq:MLL} are such that the ideal $I$ acts
  trivially, whence the action of $T(L\oplus L)$ induces an action of
  $UL(L)$.  Conversely, restricting the action of $UL(L)$ to $L\oplus L$
  gives the actions of $L$ on $M$ which satisfy conditions \eqref{eq:LLM},
  \eqref{eq:LML} and \eqref{eq:MLL} by virtue of the the relations in
  the ideal $I$.
\end{proof}

We can now define a \textbf{corepresentation} of a Leibniz algebra, as
a right module of its universal enveloping algebra.  Unlike with Lie
algebras, they are now not simply given by changing the sign of the
actions because changing the sign is \emph{not} an antiautomorphism of
the universal enveloping algebra.  Instead we have the following

\begin{proposition}
  The endomorphism of $UL(L)$, defined on generators by
  \begin{equation}
    \label{eq:antiauto}
    \ell_X \mapsto -\ell_X \qquad\text{and}\qquad
    r_X \mapsto \ell_X + r_X
  \end{equation}
  extends to an antiautomorphism of $UL(L)$.
\end{proposition}

\begin{proof}
  It is just a matter of showing that the above map preserves the
  ideal $I$.  In detail, for the first relation $R_1(X,Y)$ in
  \eqref{eq:relations}, we find
  \begin{equation}
    R_1(X,Y) \mapsto - \ell_{[X,Y]} - (-\ell_Y)\otimes(-\ell_X) +
    (-\ell_X)\otimes(-\ell_Y) = - R_1(X,Y)~;
  \end{equation}
  whereas for the second relation $R_2(X,Y)$ we find
  \begin{equation}
   R_2(X,Y) \mapsto  r_{[X,Y]} + \ell_{[X,Y]} - (\ell_Y + r_Y)\otimes
   (-\ell_X) + (-\ell_X) \otimes (\ell_Y + r_Y) = R_1(X,Y) + R_2(X,Y)~,
  \end{equation}
  and for the last relation $R_3(X,Y)$ we find
  \begin{equation}
    R_3(X,Y) \mapsto (-\ell_X + \ell_X + r_X)\otimes (\ell_Y + r_Y) =
    R_3(Y,X)~.
  \end{equation}
  One can check that restricted to the ideal $I$, this
  antiautomorphism has order 2.
\end{proof}


Unlike Lie algebras, for which homology and cohomology both can take
values in representations, for a Leibniz algebra homology takes values
in a corepresentation and cohomology takes values in a representation,
somewhat confusingly.  We will not use homology in this paper, only
cohomology with values in a representation, to which we now turn.

\subsection{Cohomology}
\label{sec:cohomology}

Let $L$ be a Leibniz algebra and $M$ a representation.  On the graded
space $CL^\bullet(L;M) = \bigoplus_{p\geq 0} CL^p(L;M)$, with
\begin{equation}
  \label{eq:complex}
  CL^p(L;M) := \Hom(L^{\otimes p},M)
\end{equation}
the space of $p$-linear maps from $L$ to $M$, we define a differential
$d: CL^p(L;M) \to CL^{p+1}(L;M)$ by the following formula
\begin{multline}
  \label{eq:differential}
  (d \varphi)(X_1, \dots, X_{p+1}) = \sum_{i=1}^p
  (-1)^{i-1} [X_i, \varphi(X_1, \dots, \widehat{X_i}, \dots, X_{p+1})]\\
  + (-1)^{p+1} [\varphi(X_1, \dots, X_p), X_{p+1}]\\
  + \sum_{1\leq i < j \leq p+1} (-1)^i \varphi( X_1, \dots
 , \widehat{X_i}, \dots, [X_i,X_j], \dots
 , X_{p+1})~,
\end{multline}
where $\varphi\in CL^p(L;M)$, $X_i \in L$ and where $\widehat{~}$
denotes omission.  Notice that we have both the left and right actions
of $L$ on $M$ in the expression for the differential.  It follows from
the explicit expression of the differential that if $N\subset M$ is a
subrepresentation, then $CL^\bullet(L;N)$ is a subcomplex.

In the special case of $M$ being a symmetric representation, the
differential takes a somewhat simpler form
\begin{multline}
  \label{eq:diffsymmetric}
  (d \varphi)(X_1, \dots, X_{p+1}) = \sum_{i=1}^{p+1}
  (-1)^{i-1} [X_i, \varphi(X_1, \dots, \widehat{X_i}
 , \dots, X_{p+1})] \\
  + \sum_{1\leq i < j \leq p+1} (-1)^i \varphi( X_1, \dots
 , \widehat{X_i}, \dots, [X_i,X_j], \dots
 , X_{p+1})~,
\end{multline}
reminiscent of the Chevalley--Eilenberg differential  computing Lie
algebra cohomology.

Let us write the first few differentials in the general case.  If $m
\in M = CL^0(L;M)$, then
\begin{equation}
  \label{eq:d0}
  dm(X) = - [m,X]~.
\end{equation}
If $\theta : L \to M$, then $d\theta: L \otimes L \to M$ is defined by
\begin{equation}
  \label{eq:d1}
  d\theta(X, Y) = [X,\theta(Y)] + [\theta(X),Y] -
  \theta([X,Y])~.
\end{equation}
Finally, if $\varphi: L \otimes L \to M$, then $d\varphi : L^{\otimes
  3} \to M$ is defined by
\begin{multline}
  \label{eq:d2}
  d\varphi(X, Y, Z) = [X,\varphi(Y, Z)] -
  [Y,\varphi(X, Z)] - [\varphi(X, Y), Z] \\
  + \varphi(X, [Y,Z]) - \varphi(Y, [X,Z]) - 
  \varphi([X,Y], Z)~.
\end{multline}
One can check that $d²=0$ precisely because $L$ is a Leibniz
algebra and $M$ is a representation \cite{MR1213376}.  Let us
illustrate this for $p=0$ and $p=1$.  Indeed, if $m\in M = CL^0(L;M)$,
then
\begin{align*}
  d^2m(X,Y) &= [X,dm(Y)] + [dm(X),Y] - dm([X,Y]) && \text{by equation
    \eqref{eq:d1}}\\
  &= - [X, [m,Y]] - [[m,X],Y] + [m,[X,Y]] && \text{by equation
    \eqref{eq:d0}}\\
  &= 0~. && \text{by equation \eqref{eq:MLL}}
\end{align*}
Similarly, if $\theta \in CL^1(L;M)$, then
\begin{align*}
  d^2\theta(X,Y,Z) &= [X,d\theta(Y, Z)] -
  [Y,d\theta(X, Z)] - [d\theta(X, Y), Z] \\
  & \quad + d\theta(X, [Y,Z]) - d\theta(Y, [X,Z]) -  d\theta([X,Y],
  Z) && \text{by equation \eqref{eq:d2}}\\
  &= [X,[Y,\theta(Z)] + [\theta(Y),Z] - \theta([Y,Z])]\\
  &\quad - [Y, [X,\theta(Z)] + [\theta(X),Z] - \theta([X,Z])]\\
  &\quad - [[X,\theta(Y)] + [\theta(X),Y] - \theta([X,Y]),Z]\\
  &\quad + [X,\theta([Y,Z])] + [\theta(X),[Y,Z]] - \theta([X,[Y,Z]])\\
  &\quad - [Y,\theta([X,Z])] - [\theta(Y),[X,Z]] + \theta([Y,[X,Z]])\\
  &\quad - [[X,Y],\theta(Z)] - [\theta([X,Y]),Z] + \theta([[X,Y],Z])
  && \text{by equation \eqref{eq:d1}}\\
\intertext{which rearranges into}
  &= \theta\left([[X,Y],Z] - [X,[Y,Z]] + [Y,[X,Z]] \right)\\
  &\quad + [X,[Y,\theta(Z)]] - [Y,[X,\theta(Z)]] - [[X,Y],\theta(Z)]\\
  &\quad + [X,[\theta(Y),Z]] - [[X,\theta(Y)],Z] - [\theta(Y),[X,Z]]\\
  &\quad - [Y,[\theta(X),Z]] - [[\theta(X),Y],Z] + [\theta(X),[Y,Z]]~,
\end{align*}
and the last four lines vanish because of equations \eqref{eq:LI},
\eqref{eq:LLM}, \eqref{eq:LML} and \eqref{eq:MLL}, respectively.

\section{$n$-Leibniz algebras and their associated Leibniz algebras}
\label{sec:LnA}

In this section we exhibit a functor from the category of (metric) $n$-Leibniz
algebras to the category of (metric) Leibniz algebras.

\subsection{From $n$-Leibniz algebras to Leibniz algebras}
\label{sec:functor}

Let $V$ be a (left) $n$-Leibniz algebra; that is, $V$ is a vector
space with a multilinear bracket $V^n \to V$, $(x_1,\dots,x_n) \mapsto
[x_1,\dots,x_n]$ obeying the fundamental identity
\begin{equation}
  \label{eq:nLI}
  [D(X), D(Y)] \cdot z = D(D(X)\cdot Y) \cdot z~, \qquad\text{for all
    $X,Y \in V^{\otimes (n-1)}$ and $z \in V$,}
\end{equation}
where $D: V^{\otimes (n-1)} \to \End V$ is defined on monomials by
\begin{equation}
  \label{eq:nD}
  D(x_1\otimes \dots \otimes x_{n-1}) \cdot z =
  [x_1,\dots,x_{n-1},z]~,
\end{equation}
and extended by linearity to all of $V^{\otimes (n-1)}$.  Because
equation \eqref{eq:nLI} is true for all $z$, it is true abstractly as
endomorphisms, whence it is formally identical to \eqref{eq:FI}, but
with $X,Y \in V^{\otimes (n-1)}$.  On $V^{\otimes (n-1)}$ we define
the following bracket
\begin{equation}
  \label{eq:nLB}
  [X,Y] := D(X) \cdot Y~.
\end{equation}

\begin{proposition}
  The above bracket turns $V^{\otimes (n-1)}$ into a Leibniz algebra.
\end{proposition}

\begin{proof}
  Indeed, the Leibniz identity \eqref{eq:LI} reads
  \begin{equation}
    D(X) \cdot (D(Y) \cdot Z) = D(D(X) \cdot Y) \cdot Z + D(Y) \cdot
    (D(X) \cdot Z)~,
  \end{equation}
  or equivalently
  \begin{equation}
    [D(X), D(Y)] \cdot Z = D(D(X)\cdot Y) \cdot Z~,
  \end{equation}
  which follows from the fundamental identity \eqref{eq:FI}.
\end{proof}

We will call $V^{\otimes (n-1)}$ the \textbf{Leibniz algebra
  associated to the $n$-Leibniz algebra} $V$ and from now on we will
denote it $L(V)$, or simply $L$ if $V$ is implicit.

The map $V \mapsto L(V)$ extends to a functor from the category of
$n$-Leibniz algebras to the category of Leibniz algebras, by sending a
morphism $\varphi: V \to W$ of $n$-Leibniz algebras, to $L\varphi:
L(V) \to L(W)$, defined on monomials by
\begin{equation}\label{eq:Lphi}
  L\varphi(x_1\otimes\dots \otimes x_{n-1}) = \varphi(x_1) \otimes
  \dots \otimes \varphi(x_{n-1})
\end{equation}
and extended to all of $L(V)$ by linearity.  It is clear from the
definition that if $\varphi: V \to W$ and $\psi: U \to V$ are
morphisms, then $L(\varphi \circ \psi) = L\varphi \circ L\psi$.

\begin{lemma}\label{le:functor}
  The map $L\varphi: L(V) \to L(W)$ defined in \eqref{eq:Lphi} is a
  morphism of Leibniz algebras.
\end{lemma}

\begin{proof}
  Recall that a linear map $\varphi: V \to W$ is a morphism of
  $n$-Leibniz algebras if
  \begin{equation}
    \label{eq:morphism}
    \varphi\left([x_1,\dots,x_n]_V\right) =
    [\varphi(x_1),\dots,\varphi(x_n)]_W~.
  \end{equation}
  If this is the case, then the Leibniz bracket of two monomials in
  the image of $L\varphi$ is given by
  \begin{align*}
    &[L\varphi(x_1\otimes\dots\otimes
    x_{n-1}),L\varphi(y_1\otimes\dots\otimes y_{n-1})]\\
    &=[\varphi(x_1)\otimes\dots\otimes
    \varphi(x_{n-1}),\varphi(y_1)\otimes\dots\otimes\varphi(y_{n-1})]\\
    &= D(\varphi(x_1)\otimes\dots\otimes\varphi(x_{n-1})) \cdot
    \left(\varphi(y_1)\otimes\dots\otimes\varphi(y_{n-1})\right)\\
    &= \sum_{i=1}^{n-1}
    \varphi(y_1) \otimes \dots \otimes
    [\varphi(x_1),\dots,\varphi(x_{n-1}),\varphi(y_i)]_W \otimes \dots
    \otimes \varphi(y_{n-1})\\
    &= \sum_{i=1}^{n-1}
    \varphi(y_1) \otimes \dots \otimes
    \varphi[x_1,\dots,x_{n-1},y_i]_V \otimes \dots \otimes
    \varphi(y_{n-1})\\
    &= \sum_{i=1}^{n-1}
    L\varphi \left(y_1 \otimes \dots \otimes [x_1,\dots,x_{n-1},y_i]_V
      \otimes \dots \otimes y_{n-1}\right)\\
    &= L\varphi\left(\sum_{i=1}^{n-1}
      y_1 \otimes \dots \otimes [x_1,\dots,x_{n-1},y_i]_V
      \otimes \dots \otimes y_{n-1}\right)\\
    &= L\varphi \left(D(x_1\otimes\dots\otimes x_{n-1})\cdot
      (y_1\otimes\dots\otimes y_{n-1})\right)\\
    &= L\varphi [x_1\otimes\dots\otimes x_{n-1}, y_1\otimes\dots\otimes
    y_{n-1}]~.
  \end{align*}
\end{proof}

There are special classes of $n$-Leibniz algebras defined by imposing
symmetry conditions on the $n$-bracket.  The best known are the
$n$-Lie algebras, where the bracket is alternating, but there are
others.  In the case of $n$-Lie algebras, the $n$-bracket defines a
map $D: \Lambda^{n-1}V \to \End V$ satisfying the fundamental identity
\eqref{eq:FI}.  Just as we did above for the general $n$-Leibniz
algebra, we may define now on $\Lambda^{n-1}V$ the structure of a
Leibniz algebra in such a way that $D$ is a Leibniz algebra morphism.
The formula for the bracket in $\Lambda^{n-1}V$ is formally identical
to equation \eqref{eq:nLB}, except that $X,Y \in \Lambda^{n-1}V$.
This closes because $\Lambda^{n-1}V$ is a $\fgl(V)$-submodule of
$V^{\otimes(n-1)}$, whence $D(X)\cdot Y \in \Lambda^{n-1}V$.  This
clearly generalises to other $n$-Leibniz algebras where the
$n$-bracket defines a map $D: T(V) \to \End V$, for $T(V)$ a
$\fgl(V)$-submodule of $V^{\otimes(n-1)}$ which becomes a Leibniz
algebra in its own right by virtue of equation \eqref{eq:nLB}.

Actually $D:T(V) \to \End V$ is strictly weaker than the conditions on
the bracket.  We really ought to say that the bracket maps $B(V) \to
V$, where $B(V) \subset V^{\otimes n}$ is a $\fgl(V)$-submodule.  This
then induces $D: T(V) \to \End V$, where $B(V) \subset T(V) \otimes
V$.  For example, for an $n$-Lie algebra, $B(V) = \Lambda^nV$ whereas
$T(V) = \Lambda^{n-1}V$.

\subsection{From metric $n$-Leibniz algebras to metric Leibniz algebras}
\label{sec:metric}

An important class of $n$-Leibniz algebras, due to their appearance in
a number of physical contexts, are those which possess an inner
product (here, a nondegenerate symmetric bilinear form) which is
invariant under inner derivations.  We will show that in this case,
the associated Leibniz algebra itself possesses a left-invariant inner
product.  Let us recall that an $n$-Leibniz algebra $V$ is said to be
\textbf{metric}, if $V$ admits an inner product $\left<-,-\right>$
which is invariant under inner derivations; that is, for all
$x_1,\dots,x_{n-1},y,z \in V$, one has
\begin{equation}
  \label{eq:metric}
  \left<[x_1,\dots,x_{n-1},y],z\right> +
  \left<y,[x_1,\dots,x_{n-1},z]\right> = 0~,
\end{equation}
or equivalently,
\begin{equation}
  \label{eq:metrictoo}
  \left<D(X) \cdot y,z\right> + \left<y,D(X)\cdot z\right> = 0~,
\end{equation}
for all $y,z \in V$ and $X\in V^{\otimes (n-1)}$.

\begin{proposition}
  Let $V$ be a metric $n$-Leibniz algebra and let $L(V) = V^{\otimes
    n}$ be its associated Leibniz algebra.  Then the natural inner
  product on $L(V)$, defined on monomials by
  \begin{equation}
    \label{eq:Lip}
    \left<x_1\otimes\dots\otimes x_{n-1},y_1\otimes\dots\otimes
      y_{n-1}\right> = \prod_{i=1}^{n-1} \left<x_i,y_i\right>
  \end{equation}
  and later extending linearly to all of $L(V)$, is invariant under
  left multiplication in $L(V)$; that is, for all $X,Y,Z \in L(V)$:
  \begin{equation}
    \label{eq:Linv}
    \left<[Z,X],Y\right> + \left<X,[Z,Y]\right> = 0~.
  \end{equation}
\end{proposition}

\begin{proof}
  Let $X = x_1\otimes\dots\otimes x_{n-1}$ and $Y=y_1
  \otimes\dots\otimes y_{n-1}$ and let $Z \in L(V)$.  Then the
  left-hand side of equation \eqref{eq:Linv} expands to
  \begin{align*}
    \left<[Z,X],Y\right> &= \left<[Z,x_1\otimes\dots\otimes
      x_{n-1}],y_1
      \otimes\dots\otimes y_{n-1}\right>\\
    &= \left<D(Z) \cdot (x_1\otimes\dots\otimes x_{n-1}),y_1
      \otimes\dots\otimes y_{n-1}\right>\\
    &= \sum_{i=1}^{n-1} \left<x_1\otimes\dots\otimes D(Z) x_i
      \otimes\dots\otimes x_{n-1}),y_1
      \otimes\dots\otimes y_{n-1}\right>\\
    &= \sum_{i=1}^{n-1} \left<D(Z) x_i, y_i\right>
    \prod_{\substack{j=1\\j\neq i}}^{n-1}
    \left<x_j,y_j\right> && \text{by equation \eqref{eq:Lip}}\\
    &= - \sum_{i=1}^{n-1} \left<x_i, D(Z) y_i\right>
    \prod_{\substack{j=1\\j\neq i}}^{n-1}
    \left<x_j,y_j\right> && \text{by equation \eqref{eq:metrictoo}}\\
    &= - \sum_{i=1}^{n-1}\left<x_1\otimes\dots\otimes x_{n-1},y_1
      \otimes\dots\otimes D(Z) y_i \otimes\dots\otimes
      y_{n-1}\right>\\
    &= - \left<x_1\otimes\dots\otimes x_{n-1},D(Z) \cdot (y_1
      \otimes\dots\otimes y_{n-1})\right>\\
    &= - \left<X,[Z,Y]\right>~.
  \end{align*}
\end{proof}

The assignment $V \mapsto L(V)$  is still functorial between the
categories of metric $n$-Leibniz algebras and metric Leibniz
algebras.  Indeed, we have the following

\begin{proposition}
  Let $\varphi: V \to W$ is a morphism of metric $n$-Leibniz algebras,
  so that in addition to equation \eqref{eq:morphism}, it is also an
  isometry:
  \begin{equation}\label{eq:isometry}
    \left<\varphi(x),\varphi(y)\right>_W = \left<x,y\right>_V~.
  \end{equation}
  Then $L\varphi: L(V) \to L(W)$ defined by equation \eqref{eq:Lphi}
  is a morphism of metric Leibniz algebras.
\end{proposition}

\begin{proof}
  Lemma \ref{le:functor} says that $L\varphi$ preserves the bracket,
  whence it remains to show that it is an isometry.  Let $X,Y \in
  L(V)$ be given by $x_1\otimes \dots \otimes x_{n-1}$ and $y_1
  \otimes \dots \otimes y_{n-1}$, respectively.  Then
  \begin{align*}
    \left<L\varphi(X),L\varphi(Y)\right> &= \left<\varphi(x_1) \otimes
      \varphi(x_{n-1}),\varphi(y_1) \otimes \varphi(y_{n-1})\right> \\
    &= \prod_{i=1}^{n-1} \left<\varphi(x_i),\varphi(y_i)\right>_W && \text{by equation \eqref{eq:Lip}}\\
    &= \prod_{i=1}^{n-1} \left<x_i,y_i\right>_V && \text{by equation \eqref{eq:isometry}}\\
    &= \left<x_1\otimes \dots \otimes x_{n-1}, y_1 \otimes \dots
      \otimes y_{n-1}\right> && \text{by equation \eqref{eq:Lip}}\\
    &= \left<X,Y\right>~.
  \end{align*}
\end{proof}

The above two propositions remain true for metric $n$-Lie algebras,
where now $L(V)= \Lambda^{n-1} V$ and the induced inner product takes
the standard determinantal form
\begin{equation}
  \label{eq:nLieIP}
  \left<x_1\wedge\dots\wedge x_{n-1},y_1\wedge\dots\wedge
    y_{n-1}\right> = \det \left(\left<x_i,y_j\right>\right)~,
\end{equation}
since the essential point is that the inner product on $V$ is
invariant under the action of $D(X)$ for any $X \in L(V)$.  One final
remark is that, although as shown in \cite{Lie3Algs} the image of $D$
in $\fso(V)$ is itself a metric Lie algebra, $D$ is not in general an
isometry.  This is easily illustrated by the unique simple euclidean
3-Lie algebra, discussed from the Faulkner point of view in
\cite[Example~4]{Lie3Algs} and from the Leibniz point of view below as
Example~\ref{eg:S4}.  Let us simply point out that here $V = \RR^4$
and $L(V) = \Lambda^2\RR^4$ and the Lie algebra of inner derivations
is $\fso(4)$.  The map $D: \Lambda^2\RR^4 \to \fso(4)$ is in this case
an isomorphism of Leibniz (and hence of Lie) algebras, but it is not
an isometry because whereas the natural inner product on $L(V)$ has
positive signature, the one on $\fso(4)$ has split signature.

\section{Deformations of an $n$-Leibniz algebra}
\label{sec:deformations}

In this section we reinterpret the deformation theory of $n$-Leibniz
algebras in terms of the cohomology of its associated Leibniz
algebra.

\subsection{Deformation complex}
\label{sec:deformation-complex}

Let $V$ be an $n$-Leibniz algebra with associated Leibniz algebra
$L=L(V)$.  Both the algebraic structures on $V$ and on $L(V)$ are given
by the Leibniz algebra morphism $D: L \to \fgl(V)$.  By a
\textbf{deformation} of $V$ (in the sense of Gerstenhaber) we mean an
analytic one-parameter family of $n$-Leibniz algebras on $V$ defined
by a bracket
\begin{equation}
  [x_1,\dots,x_n]_t = [x_1,\dots,x_n] + \sum_{k\geq 1} t^k
  \Phi_k(x_1,\dots,x_n) ~,
\end{equation}
where $\Phi_k: V^n \to V$ are multilinear maps.  Such a bracket gives
rise to a family of maps $D_t$ defined by
\begin{equation}
  D_t = D + \sum_{k\geq 1} t^k \varphi_k~,
\end{equation}
where $\varphi_k: L \to \End V$ is defined by
\begin{equation}
  \varphi_k(x_1 \otimes \dots \otimes x_{n-1}) \cdot y =
  \Phi_k(x_1,\dots,x_{n-1},y)~,
\end{equation}
for all $y \in V$.


Differentiating the fundamental identity \eqref{eq:nLI} for $D_t$ at
$t=0$, we obtain the following condition on $\varphi:=\varphi_1 \in
CL^1(L;\End V)$:
\begin{equation}
  \label{eq:infdef}
  [D(X),\varphi(Y)] + [\varphi(X),D(Y)] - D(\varphi(X)\cdot Y) -
  \varphi(D(X)\cdot Y) = 0~.
\end{equation}
Comparing this with equation \eqref{eq:d1}, but with $\varphi$
replacing $\theta$, we see that this can be written as $d\varphi = 0$,
provided that we define the actions of $L$ on $\End V$ as
\begin{equation}
  \label{eq:EndVrep}
  [X, \psi ] = [D(X),\psi] \qquad\text{and}\qquad
  [\psi, X] = [\psi,D(X)] - D(\psi \cdot X)~,
\end{equation}
for all $\psi \in \End V$ and where the brackets on the right-hand
sides are commutators on $\End V$.

\begin{proposition}
  With respect to the above actions, $\End V$ is a representation of
  $L(V)$.
\end{proposition}

\begin{proof}
  We need to show that the actions in \eqref{eq:EndVrep} satisfy
  the three compatibility conditions \eqref{eq:LLM},
  \eqref{eq:LML} and \eqref{eq:MLL}, making $\End V$ into a
  representation of $L$.  Indeed, equation \eqref{eq:LLM} is
  clear:
  \begin{align*}
    [[X,Y],\psi] -[X,[Y,\psi]] + [Y,[X,\psi]] &= [D([X,Y]),\psi] -
    [D(X),[D(Y),\psi]] +  [D(Y),[D(X),\psi]]\\
    &= [D([X,Y]),\psi] - [[D(X),D(Y)],\psi] = 0~,
  \end{align*}
  by virtue of the fundamental identity \eqref{eq:FI}.
  To check equations \eqref{eq:LML} and \eqref{eq:MLL}, it is enough to
  check one of them and equation \eqref{eq:Z}.  Let us check this latter
  equation:
  \begin{align*}
    [[X,\psi],Y] + [[\psi,X],Y] &= -[D(\psi\cdot X),Y]\\
    &= -[D(\psi\cdot X),D(Y)] + D(D(\psi\cdot X)\cdot Y) = 0~,
  \end{align*}
  again by virtue of the fundamental identity \eqref{eq:FI} but applied
  to $\psi\cdot X$ and $Y$.  Finally, we check equation \eqref{eq:LML}.
  Using the fundamental identity, the left-hand side expands to
  \begin{align*}
    [[X,\psi],Y] &= [[D(X),\psi],Y]\\
    &= [[D(X),\psi],D(Y)] - D([D(X),\psi]\cdot Y)\\
    &= [D(X),[\psi,D(Y)]] - [\psi,[D(X),D(Y)]] - D(D(X) \cdot \psi \cdot
    Y) + D(\psi \cdot D(X) \cdot Y) \\
    &= [D(X),[\psi, D(Y)]] - [\psi,D([X,Y])] - [D(X),D(\psi\cdot Y)] +
    D(\psi \cdot [X,Y])~,
  \end{align*}
  whereas the right-hand side expands to the same thing:
  \begin{align*}
    [X,[\psi,Y]] - [\psi,[X,Y]] &= [D(X),[\psi,Y]] - [\psi,D([X,Y])]
    + D(\psi\cdot[X,Y])\\
    &= [D(X),[\psi,D(Y)]] - [D(X),D(\psi\cdot Y)]\\
    & \qquad - [\psi,D([X,Y])] + D(\psi\cdot[X,Y])~.
  \end{align*}
\end{proof}

Notice that this representation is \emph{not} symmetric, whence it is
not induced from a representation of the associated Lie algebra
$\fg_L$.

A deformation is said to be \textbf{trivial} if it is due to the
action of a one-parameter subgroup $g_t$ of the general linear group
$\GL(V)$; that is, if
\begin{equation}
  \label{eq:trivdef}
  g_t\left([x_1,\dots,x_n]_t\right) = [g_t(x_1),\dots, g_t(x_n)]~,
\end{equation}
or equivalently
\begin{equation}
  \label{eq:trivdefD}
  g_t \circ D_t(X) = D(g_t \cdot X) \circ g_t~,
\end{equation}
for all $X \in L$.  Let $g_t(x) = x + t \gamma(x) + O(t^2)$ and
differentiate the above equation at $t=0$ to obtain
\begin{equation}
  \label{eq:coboundaryDef}
  \varphi(X) = - [\gamma, D(X)] + D(\gamma\cdot X) = - [\gamma, X]~,
\end{equation}
whence $\varphi = d\gamma$.

In other words, we have proved the following

\begin{theorem}\label{th:inf-def-n-Leibniz}
  Isomorphism classes of infinitesimal deformations of the $n$-Leibniz
  algebra $V$ are classified by $HL^1(L(V);\End V)$, with $\End V$ the
  nonsymmetric representation of $L(V)$ defined by equation
  \eqref{eq:EndVrep}.
\end{theorem}

If all deformations of $L$ are trivial, we say that the $n$-Leibniz algebra
$L$ is \textbf{rigid}.  A sufficient condition for rigidity is the
vanishing of $HL^1(L(V);\End V)$, but this is not necessary, since
infinitesimal deformations might be obstructed, as we now review.

\subsection{Obstructions}
\label{sec:obstructions}

Given an infinitesimal deformation of an $n$-Leibniz algebra, one
would like to know whether it integrates to a one-parameter
deformation.  Based on one's experience with the deformation theory of
other algebraic structures, one expects an infinite sequence of
obstructions (each one defined provided the previous one is overcome)
living in the same cohomology theory as the infinitesimal deformations
but one dimension higher.  Furthermore, one expects these obstruction
classes to be given by universal formulae using a natural graded Lie
algebra structure on the cohomology, as explained for Lie algebras by
Nijenhuis and Richardson in \cite{NijenhuisRichardson}.  We will see
that this is indeed the case in the next section, but for now let us
illustrate this by trying to integrate an infinitesimal deformation to
second order.

We write the deformed $n$-bracket on $V$ as
\begin{equation}
  [x_1,\dots,x_n]_t =   [x_1,\dots,x_n] + t \Phi_1(x_1,\dots,x_n) + t^2
  \Phi_2(x_1,\dots,x_n) + O(t^3)~,
\end{equation}
giving rise to $D_t: V^{\otimes(n-1)} \to \End V$ defined by
\begin{equation}
  D_t(X) = D(X) + t \varphi_1(X) + t^2 \varphi_2(X) + O(t^3)~,
\end{equation}
where $\varphi_i \in CL^1(L;\End V)$.  Expanding the fundamental identity
\eqref{eq:FI} for $D_t$ to order $t^2$, one finds to zeroth order the
fundamental identity for $D$, to first order the cocycle condition for
$\varphi_1$ and to second order the following identity
\begin{multline*}
  [D(X),\varphi_2(Y)] +  [\varphi_2(X),D(Y)] +
  [\varphi_1(X),\varphi_1(Y)]\\
  = D(\varphi_2(X)\cdot Y) + \varphi_2(D(X)\cdot Y) +
  \varphi_1(\varphi_1(X)\cdot Y)~,
\end{multline*}
which we recognise as
\begin{equation}
  \label{eq:obstruction2}
  d\varphi_2(X,Y) = \varphi_1(\varphi_1(X)\cdot Y) -
  [\varphi_1(X),\varphi_1(Y)]~.
\end{equation}
It is a straightforward calculation, using that $\varphi_1$ is a
cocycle, to show that the right-hand side of this equation defines a
cocycle in $CL^2(L;\End V)$ whose cohomology class is the obstruction
to integrability (to second order), since if and only if this class
vanishes, can we find $\varphi_2$ obeying equation
\eqref{eq:obstruction2}.  We will be able to interpret the right-hand
side of \eqref{eq:obstruction2} as a bracket $-\half
[\varphi_1,\varphi_1]$ in $HL^2(L;\End V)$ analogous to the
Nijenhuis--Richardson \cite{NijenhuisRichardson} bracket on the
Chevalley--Eilenberg cohomology $H^\bullet(\fg;\fg)$ of a Lie algebra
$\fg$.  This will allow us to prove in complete generality that the
obstructions to integrating an infinitesimal deformation are
cohomology classes in $HL^2(L;\End V)$.

\subsection{Another look at the deformation complex}
\label{sec:another-look}

We may understand the deformation complex in a slightly different way,
which serves to illustrate a number of things.  First of all, we
notice that a deformation of the $n$-Leibniz algebra $V$ implies a
deformation of the underlying Leibniz algebra $L(V)$.  However the
notions of trivial deformations do not agree.  A deformation of the
Leibniz algebra $L(V)$ is trivial if it is due to the action of a
one-parameter subgroup of $\GL(L(V))$, but since not every invertible
linear transformation of $L(V)$ is induced from one of $V$, we may
have that a deformation of $L(V)$ may be trivial without the
corresponding deformation of $V$ being trivial.

A deformation of the Leibniz algebra $L:=L(V)$ takes the form
\begin{equation}
  [X,Y]_t = [X,Y] + t \Psi(X,Y) + O(t^2)~,
\end{equation}
where $\Psi: L^2 \to L$ is a bilinear map.   Expanding the Leibniz identity
\eqref{eq:LLI} for the deformed bracket to first order recovers, at
zeroth order, the Leibniz identity for the undeformed bracket and, at
first order, the following equation for $\Psi$:
\begin{multline}
  \label{eq:Ldef2}
  [X,\Psi(Y,Z)] - [Y,\Psi(X,Z)] - [\Psi(X,Y),Z]\\
  + \Psi(X,[Y,Z]) - \Psi(Y,[X,Z])  - \Psi([X,Y],Z) = 0~.
\end{multline}
Comparing with the expression \eqref{eq:d2} for the differential in
Leibniz cohomology, we see that this is the cocycle condition for
$\Psi \in CL^2(L;L)$.  The deformation is trivial if it is the result
of the action of a one-parameter subgroup $g_t$ of the general linear group
$\GL(L)$, so that
\begin{equation}
  g_t\left([X,Y]_t\right) = [g_t(X),g_t(Y)]~.
\end{equation}
Letting $g_t(X) = X + t \gamma(X) + O(t^2)$ and differentiating the
above equation with respect to $t$ at $t=0$ we obtain
\begin{equation}
  \Psi(X,Y) = [X,\gamma(Y)] + [\gamma(X),Y] - \gamma([X,Y])~,
\end{equation}
whence $\Psi = d\gamma$, for $\gamma \in CL^1(L;L)$.

This proves the following

\begin{theorem}
  \label{th:infdefL}
  Infinitesimal deformations of a Leibniz algebra $L$ are classified
  by $HL^2(L;L)$.
\end{theorem}

Now we have a vector space isomorphism
\begin{equation}
  CL^{p+1}(L;L) = \Hom(L^{\otimes (p+1)},L) \cong \Hom(L^{\otimes p},\End
  L) = CL^p(L;\End L)~,
\end{equation}
for $p\geq0$.  We may promote this to an isomorphism of complexes by
defining the differential on $CL^\bullet(L;\End L)$ appropriately,
which will tell us in turn how to view $\End L$ as a representation of
$L$.

In lowest dimension, we must impose the commutativity of the following
diagram:
\begin{equation}
  \begin{CD}
    CL^1(L;L) @>{\cong}>> CL^0(L;\End L)\\
    @V{d}VV                  @VV{d}V\\
    CL^2(L;L) @>{\cong}>> CL^1(L;\End L)\\
    @V{d}VV                  @VV{d}V\\
    CL^3(L;L) @>{\cong}>> CL^2(L;\End L)~,
  \end{CD}
\end{equation}
for some suitable $d: CL^p(L;\End L) \to CL^{p+1}(L;\End L)$
determined by how $L$ acts on $\End L$.  It is this action which we
will determine.

Consider $\psi \in CL^1(L;L) = \End L$.  Then for $\psi \in
CL^1(L;L)$, $d\psi \in CL^2(L;L)$ is given by
\begin{equation}
  d\psi(X,Y) = [X,\psi(Y)] + [\psi(X),Y] - \psi([X,Y])~.
\end{equation}
On the other hand, for $\psi \in CL^0(L;\End L) = \End L$,
\begin{equation}
  d\psi(X) = -[\psi, X]~,
\end{equation}
whence demanding commutativity of the top square,
\begin{equation}
  d\psi(X)(Y) = -[\psi,X](Y) = [X,\psi(Y)] + [\psi(X),Y] -
  \psi([X,Y])~,
\end{equation}
which says that the right action of $L$ on $\End L$ is given by
\begin{equation}
  \label{eq:EndL-L}
  [\psi,X](Y) = - [X,\psi(Y)] - [\psi(X),Y] + \psi([X,Y])~.
\end{equation}

Now let $\Phi \in CL^2(L;L)$ and let the corresponding element in
$CL^1(L;\End L)$ be $\varphi$; that is, $\varphi(X)(Y) = \Phi(X,Y)$.
Then on the one hand,
\begin{multline*}
  d\Phi(X,Y,Z) = 
  [X,\Phi(Y,Z)] - [Y,\Phi(X,Z)] - [\Phi(X,Y),Z]\\
  + \Phi(X,[Y,Z]) - \Phi(Y,[X,Z])  - \Phi([X,Y],Z)~,
\end{multline*}
which we would like to equate with
\begin{equation}
  d\varphi(X,Y) = [X,\varphi(Y)] + [\varphi(X),Y] - \varphi([X,Y])
\end{equation}
applied to $Z$:
\begin{multline*}
  d\varphi(X,Y)(Z) = [X,\varphi(Y)](Z)  - [Y,\Phi(X,Z)] -
  [\Phi(X,Y),Z] + \Phi(X,[Y,Z]) - \Phi([X,Y],Z)~,
\end{multline*}
where we have used equation \eqref{eq:EndL-L}.  Comparing the two
expressions determines the left action of $L$ on $\End L$ to be
\begin{equation}
  \label{eq:L-EndL}
  [X,\psi](Y) = [X,\psi(Y)] - \psi([X,Y])~.
\end{equation}

\begin{proposition}
  With the actions defined by \eqref{eq:EndL-L} and \eqref{eq:L-EndL},
  $\End L$ is a representation of $L$.
\end{proposition}

\begin{proof}
  We need to check that conditions \eqref{eq:LLM}, \eqref{eq:LML} and
  \eqref{eq:MLL} are satisfied.

  Checking condition \eqref{eq:LLM} we apply it to $Z\in L$ and use
  the Leibniz identity for $L$ to expand the left-hand side as
  follows:
  \begin{align*}
    [[X,Y],\psi](Z) &= [[X,Y],\psi(Z)] - \psi([[X,Y],Z])\\
    &= [X,[Y,\psi(Z)]] - [Y,[X,\psi(Z)]] - \psi([X,[Y,Z]]) +
    \psi([Y,[X,Z]])~,
  \end{align*}
  whereas expanding the right-hand side we obtain, for the first term
  \begin{align*}
    [X,[Y,\psi]](Z) &= [X,[Y,\psi](Z)] - [X,\psi]([Y,Z])\\
    &= [X,[Y,\psi(Z)]] - [X,\psi([Y,Z])] - [Y,\psi([X,Z])] +
    \psi([Y,[X,Z]])~,
  \end{align*}
  and for the second term
  \begin{align*}
    - [Y,[X,\psi]](Z) &= -[Y,[X,\psi](Z)] + [Y,\psi]([X,Z])\\
    &= -[Y,[X,\psi(Z)]] + [Y,\psi([X,Z])] + [X,\psi([Y,Z])] -
    \psi([X,[Y,Z]])~.
  \end{align*}
  Adding them we find that four of the terms cancel pairwise and the
  remaining four are precisely what we obtained for the left-hand
  side.

  In order to check equations \eqref{eq:LML} and \eqref{eq:MLL}, it is
  enough to check one of them and equation \eqref{eq:Z}.  Applying
  this latter equation to $Z \in L$ and expanding, we obtain for the
  first term,
  \begin{align*}
    [[X,\psi],Y](Z) &= - [Y, [X,\psi](Z)] - [[X,\psi](Y),Z] +
    [X,\psi]([Y,Z])\\
    &= - [Y,[X,\psi(Z)]] + [Y,\psi([X,Z])] - [[X,\psi(Y)],Z]\\
    & \qquad + [\psi([X,Y]),Z] + [X,\psi([Y,Z])] - \psi([X,[Y,Z]])~,
  \end{align*}
  and for the second
  \begin{align*}
    [[\psi,X],Y](Z) &= - [[Y, [\psi,X](Z)] - [[\psi,X](Y),Z] +
    [\psi,X]([Y,Z])\\
    &=  [Y, [X,\psi(Z)]] + [Y,[\psi(X),Z]] - [Y,\psi([X,Z])]\\
    &\qquad  +  [[X,\psi(Y)],Z] + [[\psi(X),Y],Z] - [\psi([X,Y]),Z]\\
    &\qquad - [X,\psi([Y,Z])] - [\psi(X),[Y,Z]] + \psi([X,[Y,Z]])~.
  \end{align*}
  Adding the two, six terms cancel pairwise, leaving
  \begin{equation}
    [[X,\psi],Y](Z)  + [[\psi,X],Y](Z) = [Y,[\psi(X),Z]]  +
    [[\psi(X),Y],Z] - [\psi(X),[Y,Z]]~,
  \end{equation}
  which vanishes because of the Leibniz identity \eqref{eq:LI} with
  $\psi(X)$ replacing $X$.  Finally, we check equation \eqref{eq:LML},
  by applying it to $Z$ and expanding.  Doing so with the left-hand
  side we find
  \begin{align*}
    [[X,\psi],Y](Z) &= -[Y,[X,\psi](Z)] - [[X,\psi](Y),Z] +
    [X,\psi]([Y,Z])\\
    &= - [Y, [X,\psi(Z)]] + [Y,\psi([X,Z])] - [[X,\psi(Y)],Z]\\
    & \qquad + [\psi([X,Y]),Z] + [X,\psi([Y,Z])] - \psi([X,[Y,Z]])~,
  \end{align*}
  whereas for the right-hand side we find
  \begin{align*}
    [X,[\psi,Y]](Z) - [\psi,[X,Y]](Z) &= [X,[\psi,Y](Z)] -
    [\psi,Y]([X,Z]) + [[X,Y],\psi(Z)]\\
    &\qquad + [\psi([X,Y]),Z] - \psi([[X,Y],Z])\\
    &= - [X,[Y,\psi(Z)]] - [X,[\psi(Y),Z]] + [X,\psi([Y,Z])]\\
    &\qquad + [Y,\psi([X,Z])] + [\psi(Y),[X,Z]] - \psi([Y,[X,Z]])\\
    &\qquad + [[X,Y],\psi(Z)] + [\psi([X,Y]),Z] - \psi([[X,Y],Z])~.
  \end{align*}
  Comparing the two we find that eight terms cancel pairwise in their
  difference and the rest are
  \begin{multline*}
    [Y,[X,\psi(Z)]] - [X,[Y,\psi(Z)]] + [[X,Y],\psi(Z)] \\
    + [[X,\psi(Y)],Z] - [X,[\psi(Y),Z]] + [\psi(Y),[X,Z]]\\
    + \psi([X,[Y,Z]]) - \psi([Y,[X,Z]]) - \psi([[X,Y],Z])~,
  \end{multline*}
  each line of which vanishes because of the Leibniz identity.
\end{proof}


We therefore have two complexes $CL^{\bullet + 1}(L;L)$ and
$CL^\bullet(L;\End L)$ which are isomorphic as graded vector spaces
and, as seen above, also isomorphic as complexes in the lowest two
degrees.  In fact, we have more.

\begin{proposition}
  The vector space isomorphism $CL^p(L;\End
  L) \to CL^{p+1}(L;L)$, sending $\varphi \mapsto \Phi$, where
  \begin{equation}
    \label{eq:complexiso}
    \varphi(X_1,\dots,X_p)(Y) = \Phi(X_1,\dots,X_p,Y)~,
  \end{equation}
  is an isomorphism of complexes.
\end{proposition}

\begin{proof}
  We need to show that the map defined by \eqref{eq:complexiso} is a
  chain map.  By equation \eqref{eq:differential}, applying
  $(d\varphi)(X_1,\dots,X_{p+1})$ to $X_{p+2}\in L$, we obtain
  \begin{align*}
    (d\varphi)(X_1,\dots,X_{p+1})(X_{p+2}) &= \sum_{i=1}^p
    (-1)^{i-1} [X_i, \varphi(X_1, \dots, \widehat{X_i}, \dots, X_{p+1})](X_{p+2})\\
    & \quad + (-1)^{p+1} [\varphi(X_1, \dots, X_p), X_{p+1}](X_{p+2})\\
    & \quad + \sum_{1\leq i < j \leq p+1} (-1)^i \varphi( X_1, \dots
    , \widehat{X_i}, \dots, [X_i,X_j], \dots, X_{p+1})(X_{p+2})~,\\
    \intertext{which becomes, using equations \eqref{eq:EndL-L} and
      \eqref{eq:L-EndL},}
    &= \sum_{i=1}^p (-1)^{i-1} [X_i, \Phi(X_1, \dots,
    \widehat{X_i}, \dots, X_{p+1},X_{p+2})]\\
    &\quad - \sum_{i=1}^p (-1)^{i-1} \Phi(X_1,\dots,
    \widehat{X_i}, \dots, X_{p+1},[X_i,X_{p+2}]) \\
    & \quad - (-1)^{p+1}  [X_{p+1}, \Phi(X_1, \dots, X_p,X_{p+2})]\\
    &\quad - (-1)^{p+1} [\Phi(X_1,\dots,X_{p+1}),X_{p+2}]\\
    &\quad + (-1)^{p+1} \Phi(X_1,\dots, X_p, [X_{p+1},X_{p+2}]) \\
    & \quad + \sum_{1\leq i < j \leq p+1} (-1)^i \Phi( X_1, \dots
    , \widehat{X_i}, \dots, [X_i,X_j], \dots, X_{p+1},X_{p+2})~.
  \end{align*}
  The first and third terms make up
  \begin{equation}
    \sum_{i=1}^{p+1} (-1)^{i-1} [X_i, \Phi(X_1, \dots, \widehat{X_i},
    \dots, X_{p+2})]~,
  \end{equation}
  whereas the second term and the last two make up
  \begin{equation}
    \sum_{1\leq i < j \leq p+2} (-1)^i \Phi( X_1, \dots
    , \widehat{X_i}, \dots, [X_i,X_j], \dots, X_{p+2})~.
  \end{equation}
  Putting everything together we arrive at
  \begin{align*}
    (d\varphi)(X_1,\dots,X_{p+1})(X_{p+2}) &= 
    \sum_{i=1}^{p+1} (-1)^{i-1} [X_i, \Phi(X_1, \dots, \widehat{X_i}, \dots, X_{p+2})]\\
    & \quad + (-1)^{p+2} [\Phi(X_1, \dots, X_{p+1}), X_{p+2}]\\
    & \quad + \sum_{1\leq i < j \leq p+2} (-1)^i \Phi(X_1, \dots
    , \widehat{X_i}, \dots, [X_i,X_j], \dots, X_{p+2})\\
    &= d\Phi(X_1,\dots,X_{p+2})~,
  \end{align*}
  using equation \eqref{eq:differential}, whence the isomorphism
  \eqref{eq:complexiso} is a chain map and hence an isomorphism of
  complexes.
\end{proof}

Together with Theorem~\ref{th:infdefL}, the above isomorphism of
complexes implies the following

\begin{theorem}
  Infinitesimal deformations of a Leibniz algebra $L$ are classified
  by $HL^1(L;\End L)$ with $\End L$ the representation of $L$ defined
  by \eqref{eq:EndL-L} and \eqref{eq:L-EndL}.
\end{theorem}

Back to the deformations of an $n$-Leibniz algebra $V$, since $L(V) =
V^{\otimes (n-1)}$ is a faithful $\GL(V)$-module, we have an injective
map $\iota: \End V \to \End L(V)$.  If $\psi\in\End V$, its image
$\iota(\psi) \in \End L(V)$ is such that if $x_1\otimes \dots\otimes
x_{n-1} \in L(V)$ is a monomial,
\begin{equation}
  \iota(\psi)(x_1\otimes \dots\otimes x_{n-1}) = \sum_{i=1}^{n-1} x_1
  \otimes \dots \otimes \psi(x_i) \otimes \dots \otimes x_{n-1}~,
\end{equation}
and we extend to all of $L(V)$ by linearity.  We will often use the
shorthand notation $\psi\cdot X$ for $\iota(\psi)(X)$, as was done in
\eqref{eq:Leib1}, for instance.

\begin{proposition}
  The map $\iota: \End V \to \End L(V)$ is a morphism of $L(V)$
  representations, where $L(V)$ acts on $\End V$ according to
  \eqref{eq:EndVrep}.
\end{proposition}

\begin{proof}
  We want to show that, for all $X \in L(V)$ and $\psi \in \End V$,
  the following relations hold:
  \begin{equation}
    [X,\iota(\psi)] = \iota \left([X,\psi]\right)\qquad\text{and}\qquad
    [\iota(\psi),X] = \iota \left([\psi,X]\right)~.
  \end{equation}
  For $Y \in L(V)$ we have
  \begin{align*}
    [X,\iota(\psi)](Y) &= [X, \psi\cdot Y] - \psi \cdot [X,Y] &&
    \text{by equation \eqref{eq:L-EndL}}\\
    &= D(X) \cdot \psi\cdot Y - \psi \cdot D(X)\cdot Y\\
    &= [D(X),\psi]\cdot Y\\
    &= \iota([X,\psi])(Y) && \text{by the first equation in
      \eqref{eq:EndVrep}.}
  \end{align*}
  Similarly,
  \begin{align*}
    [\iota(\psi),X](Y) &= -[X,\iota(\psi)(Y)] - [\iota(\psi)(X),Y] +
    \iota(\psi)([X,Y]) && \text{by equation \eqref{eq:EndL-L}}\\
    &= -D(X)\cdot \psi \cdot Y - D(\psi\cdot X)\cdot Y +
    \psi \cdot D(X) \cdot Y\\
    &= [\psi,D(X)]\cdot Y  - D(\psi \cdot X) \cdot Y\\
    &= [\psi,X]\cdot Y && \text{by the second equation in
      \eqref{eq:EndVrep}}\\
    &= \iota([\psi,X])(Y)~.
  \end{align*}
\end{proof}

This map induces an injective map of complexes $CL^\bullet(L;\End V)
\to CL^\bullet(L;\End L)$, whence the deformation complex for the
$n$-Leibniz algebra written in Section~\ref{sec:deformation-complex}
is a subcomplex of the deformation complex of the associated Leibniz
algebra $L(V)$.  It is preferable, however, to work with
$CL^\bullet(L;\End V)$ itself.

\subsection{A graded Lie algebra structure on the deformation complex}
\label{sec:graded-lie-algebra}

In the study of deformations of Lie algebras, many calculations become
simpler by first exhibiting a graded Lie algebra structure on the
deformation complex, relative to which the differential is an inner
derivation.  As a consequence, the cocycles are a (graded Lie)
subalgebra of which the coboundaries are an ideal, whence the
cohomology itself inherits the structure of a graded Lie algebra.  The
same situation obtains in the deformation theory of $n$-Leibniz
algebras.

Let us depart from the observation that if $V$ is an $n$-Leibniz
algebra with associated Leibniz algebra $L$, then $D: L \to \End V$
can be understood as a cochain $D \in CL^1(L;\End V)$.  This cochain
is actually a cocycle:
\begin{align*}
  dD(X, Y) &= [X,D(Y)] + [D(X),Y] - D([X,Y])\\
  &= [D(X),D(Y)] + [D(X),D(Y)] - D(D(X)\cdot Y) - D([X,Y])\\
  &= 2 ([D(X),D(Y)] -D([X,Y])) = 0~,
\end{align*}
by the fundamental identity.  Furthermore $D$ is a coboundary.
Indeed, let $\1 \in \End V$ denote the identity endomorphism and
consider
\begin{equation}
  d\1(X) = - [\1,X] = -[\1,D(X)] + D(\1\cdot X) = (n-1) D(X)~,
\end{equation}
whence $D = \frac1{n-1} d\1$.

If we define, for $\alpha,\beta \in CL^1(L;\End V)$, their bracket
$[\alpha,\beta] \in CL^2(L;\End V)$ by
\begin{equation}\label{eq:GLBEndV}
  [\alpha,\beta](X,Y) = [\alpha(X),\beta(Y)] - \alpha(\beta(X)\cdot Y)
  + [\beta(X),\alpha(Y)] - \beta(\alpha(X)\cdot Y)~,
\end{equation}
then we have that $[D,\beta] = d\beta$ and that $[D,D]=0$ because of
the fundamental identity, whence $d^2=0$.  This suggests very strongly
the following: the bracket above extends to a graded Lie bracket on
all of $CL^\bullet(L;\End V)$ and the differential in the complex is
given by $[D,-]$.  This turns out to be the case and the existence
of this graded Lie algebra structure on $CL^\bullet(L;\End V)$ can be
deduced in at least two ways:
\begin{enumerate}
\item from work of Balavoine \cite{MR1436922} for Leibniz algebras,
  via the maps
  \begin{equation} CL^\bullet(L;\End V) \into CL^\bullet(L;\End L)
    \cong CL^{\bullet+1}(L;L)~.
  \end{equation}
  The latter map pulls back to $CL^\bullet(L;\End L)$
  the graded Lie algebra structure on $CL^{\bullet+1}(L;L)$ defined by
  Balavoine,  and one checks that $CL^\bullet(L;\End V)$ is a graded
  Lie subalgebra; or
\item more directly from work of Rotkiewicz \cite{MR2243339} who
  defines a graded Lie algebra structure for a cohomology complex
  associated to an $n$-Leibniz algebra, which is isomorphic to
  $CL^\bullet(L;\End V)$.
\end{enumerate}

In either case, we have the following

\begin{theorem}[Balavoine \cite{MR1436922}, Rotkiewicz \cite{MR2243339}]
  The complex $CL^\bullet(L;\End V)$ admits the structure of a graded
  Lie algebra in such a way that the differential is an inner
  derivation $d = [D,-]$ by an element $D \in CL^1(L;\End V)$ obeying
  $[D,D]=0$.
\end{theorem}

In what follows we will not need the explicit expression for the
graded Lie bracket; it will be enough to know that it exists and that
in the expression $[\alpha,\beta](X_1,\dots,X_{p+q})$ where $\alpha
\in CL^p(L;\End V)$ and $\beta \in CL^q(L;\End V)$ and $X_i \in L$,
there are only two kinds of terms: commutators of the form
$[\alpha(X_{i_1},\dots,X_{i_p}),\beta(X_{i_{p+1}},\dots,X_{i_{p+q}})]$
in $\End V$ and terms in the image of either $\alpha$ or $\beta$ as in
the example in equation \eqref{eq:GLBEndV}.


Since the fundamental identity is equivalent to $[D,D]=0$, we may
consider the fundamental identity of the deformation $[D_t, D_t] =0$
and expand it in powers of $t$.  Let us assume that we have a 
deformation to order $t^N$.  This means that we have
\begin{equation}
  D_N = D + \sum_{k=1}^N t^k \varphi_k
\end{equation}
satisfying
\begin{equation}
  [D_N,D_N] = t^{N+1} \xi + O(t^{N+2})~.
\end{equation}
We claim that $\xi$ is a cocycle.  This is equivalent to showing that
$[D,[D_N,D_N]] = O(t^{N+2})$, but this is clear because
\begin{equation}
  [D,[D_N,D_N]] = [D_N - (D_N-D), [D_N,D_N]] =
  [D_N,[D_N,D_N]] - [D_N-D,[D_N,D_N]]~,
\end{equation}
and the first term in the right-hand side vanishes because of the
Jacobi identity and, since $D_N - D = O(t)$, the second term is
$O(t^{N+2})$ as desired.

Furthermore if and only if the class of $\xi$ in $HL^2(L;\End V)$
vanishes, so that $\xi = -2 d\varphi_{N+1}$, can we extend the
deformation to the next order by defining
\begin{equation}
  D_{N+1} := D_N + t^{N+1} \varphi_{N+1}
\end{equation}
and noticing that now $[D_{N+1},D_{N+1}] = O(t^{N+2})$.
In this way we arrive at an infinite sequence of obstructions in
$HL^2(L;\End V)$ for integrating an infinitesimal deformation.

In summary,  we have the following more complete version of Theorem
\ref{th:inf-def-n-Leibniz}.

\begin{theorem}\label{th:def-n-Leibniz}
  Infinitesimal deformations of an $n$-Leibniz algebra $V$ are
  classified by $HL^1(L;\End V)$ with $\End V$ the representation of
  $L=V^{\otimes (n-1)}$ defined by \eqref{eq:EndVrep}.  The
  obstructions to integrating an infinitesimal deformation live in
  $HL^2(L;\End V)$.
\end{theorem}

It follows that if $HL^1(L;\End V) = 0$, the $n$-Leibniz algebra $L$
is rigid, whereas infinitesimal deformations are unobstructed
if $HL^2(L;\End V) = 0$.  Of course, even if $HL^1(L;\End V) \neq 0$,
an infinitesimal deformation may fail to integrate and $L$ may still
be rigid.  Similarly, even if $HL^2(L;\End V) \neq 0$, infinitesimal
deformations may still be unobstructed.  We end with the remark that
because $\End V$ is not a symmetric representation, one cannot use Lie
algebra cohomology to compute $HL^\bullet(L;\End V)$, whence it seems
that these groups have to be computed using brute force.

\subsection{Deformations of $n$-Lie algebras}
\label{sec:n-Lie}

Now we consider the case of $n$-Lie algebras, where the bracket is
totally skewsymmetric and hence the associated Leibniz algebra is $L =
\Lambda^{n-1}V$.  It is clear that the preceding discussion applies
\emph{mutatis mutandis} (which here simply means replacing
$\Lambda^{n-1}V$ for $V^{\otimes (n-1)}$ everywhere), except for one
important difference:
not every cocycle in $CL^1(L;\End V)$ gives rise to deformation of the
$n$-Lie algebra: we still have to impose that the resulting
$n$-bracket be totally skewsymmetric, for whereas every totally
skewsymmetric $n$-bracket defines a map $\Lambda^{n-1}V \to \End V$,
the converse does not hold: a map $\Lambda^{n-1}V \to \End V$ defines
an $n$-bracket $\Lambda^{n-1}V \otimes V \to V$, which may or may not
be skewsymmetric.  We may circumvent this problem by defining a
subcomplex $C^\bullet$ of $CL^\bullet(L;\End V)$ which agrees with
$CL^\bullet(L;\End V)$ for $p\neq 1$ and such that $C^1 \subsetneq
CL^1(L;\End V)$ consists of those $\varphi: L \to \End V$ such that
the associated $n$-linear map
\begin{equation}\label{eq:assoc-n-map}
  \Phi(x_1,\dots,x_n):=\varphi(x_1,\dots,x_{n-1})(x_n)
\end{equation}
is totally skewsymmetric.

\begin{lemma}\label{le:subcomplex}
  The subspace $C^\bullet$ so defined is a subcomplex of
  $CL^\bullet(L;\End V)$.
\end{lemma}

\begin{proof}
  We need only verify that the image of the differential $d:
  CL^0(L;\End V) \to CL^1(L;\End V)$ actually lives inside $C^1$.
  To this end let $f \in \End V = CL^0(L;\End V)$.  Let $X = x_1
  \wedge \dots \wedge x_{n-1} \in L$ and consider the $n$-linear map
  associated with $df \in CL^1(L;\End V)$:
  \begin{align*}
    df(X)(x_n) &= -[f,X](x_n)\\
    &= - [f, D(X)](x_n) +
    D(f\cdot X)(x_n)\\
    &= - f(D(X)\cdot x_n) + D(X)\cdot f(x_n) + D(f\cdot
    X)(x_n)\\
    &= -f([x_1,\dots,x_n]) + [x_1,\dots,x_{n-1},f(x_n)] +
    \sum_{i=1}^{n-1} [x_1,\dots,f(x_i),\dots,x_n]\\
    &= - -f([x_1,\dots,x_n]) + \sum_{i=1}^n
    [x_1,\dots,f(x_i),\dots,x_n]~,
  \end{align*}
  which is clearly skewsymmetric in the $x_i$.
\end{proof}

As a corollary of the proof of the previous lemma, we see that if the
$n$-bracket is such that it maps $B(V) \to V$, for some
$\fgl(V)$-submodule $B(V) \subset V^{\otimes n}$, then $df \in
CL^1(L;\End V)$ will be such that its associated $n$-linear map also
maps $B(V) \to V$.  This follows because the calculation in the proof
shows that
\begin{equation}
  df(X)(x_n) = \Phi(f \cdot (X \otimes x_n)) - f(\Phi(X
  \otimes x_n)) = -(f \cdot \Phi)(X \otimes x_n)~,
\end{equation}
where $\Phi$ stands for the $n$-bracket in $V$ and where $f \in
\End V$ acts in the natural way on all the objects.  In other words,
since $df = - f \cdot \Phi$, it is clear that $df$ will have the same
symmetries as $\Phi$.

For the case of the $n$-Lie algebras, we can therefore conclude the
following:

\begin{theorem}\label{th:def-n-Lie}
  Infinitesimal deformations of an $n$-Lie algebra are classified by
  $H^1(C^\bullet)$, where $C^\bullet \subset CL^\bullet(L;\End V)$ is
  the subcomplex defined above Lemma \ref{le:subcomplex}.  The
  obstructions to integrating an infinitesimal deformations live in
  $H^2(C^\bullet)$.
\end{theorem}

We remark that whereas the natural map $H^1(C^\bullet) \to H^1(L;\End
V)$ is injective, the surjection $H^2(C^\bullet) \to HL^2(L;\End V)$
may have kernel.  Hence whereas $HL^1(L;\End V)=0$ is a sufficient
(but not necessary) condition for $V$ to be infinitesimally rigid,
$HL^2(L;\End V)=0$ does not imply that infinitesimal deformations are
unobstructed.  In practice and in the absence of any strong structural
results, these calculations are done by explicitly solving the cocycle
and coboundary conditions in $C^\bullet$, whence the above subtleties
will not play any rôle.

\subsection{Deformations of metric $n$-Leibniz algebras}
\label{sec:metric-deformations}

Many of the more physically interesting $n$-Leibniz algebras are
metric --- a concept defined in Section~\ref{sec:metric} --- and in
setting up a deformation theory one might wish to restrict the
deformations to the class of metric $n$-Leibniz algebras.  The data
defining a metric $n$-Leibniz algebra consists of a vector space $V$
with two additional structures, the $n$-bracket and the inner product,
satisfying an open condition, namely the nondegeneracy of the inner
product, and two algebraic conditions, namely the fundamental identity
\eqref{eq:nLI} and the compatibility condition \eqref{eq:metric} with
the inner product.

A first, perhaps naive, approach to the deformation problem would be
to deform both the $n$-bracket and the inner product.  However we
notice that we can always undo the deformation of the inner product
via a change of basis: this is simply the fact that we can bring any
inner product to a diagonal normal form with entries $\pm 1$.  So it
may be better to fix the inner product once and for all and deform the
bracket in such a way that the compatibility condition
\eqref{eq:metric} is preserved.  That compatibility condition is
equivalent to \eqref{eq:metrictoo}, which says that $D: L(V) \to
\fso(V)$, where by $\fso(V)$ we mean the Lie algebra of skewsymmetric
endomorphisms of $V$.  This suggests restricting the deformation
complex to a subspace $CL^\bullet(L;\fso(V)) \subset CL^\bullet(L;\End
V)$.  To see that this is not obviously wrong, notice that an
infinitesimal metric deformation $\varphi : L \to \fso(V)$ is trivial
if $\varphi = df$, where $f \in \fso(V)$, so that we are only allowed
an orthogonal change of basis.  We have something to check, though.

\begin{proposition}\label{pr:metric-subcomplex}
  $CL^\bullet(L;\fso(V))$ is a subcomplex of $CL^\bullet(L;\End V)$.
\end{proposition}

\begin{proof}
  This follows from the fact that $\fso(V)$ is an
  $L$-subrepresentation of $\End V$.  The action of $L$ on $\End V$ is
  given by equation \eqref{eq:EndVrep}.  If $V$ is a metric
  $n$-Leibniz algebra, then $D(X) \in \fso(V)$ for all $X \in L$,
  whence if $\psi \in \fso(V)$, so are $[X,\psi] = [D(X),\psi]$, since
  $\fso(V)$ is a Lie subalgebra, and $[\psi,X] = [\psi, D(X)] - D(\psi
  \cdot X)$, for the same reasons.
\end{proof}

A refinement of this result is the following

\begin{proposition}\label{pr:metric-subalgebra}
  $CL^\bullet(L;\fso(V))$ is a graded Lie subalgebra of
  $CL^\bullet(L;\End V)$.
\end{proposition}

\begin{proof}
  This is clear from the explicit expression given in \cite{MR2243339}
  for the Lie bracket $[\alpha,\beta]$, where $\alpha \in
  CL^p(L;\fso(V))$ and $\beta \in CL^q(L;\fso(V))$.  Applying this to
  $X_1\otimes \dots \otimes X_{p+q} \in L^{\otimes (p+q)}$, we see
  that it involves two kinds of terms: commutators in $\fso(V)$ or
  terms in the image of either $\alpha$ or $\beta$ and hence both lie
  again in $\fso(V)$.  We can see this explicitly in expression
  \eqref{eq:GLBEndV} for the case $p=q=1$.
\end{proof}

We remark that Proposition \ref{pr:metric-subalgebra} implies
Proposition \ref{pr:metric-subcomplex}, since the differential $d =
[D,-]$ is the inner derivation defined by $D \in CL^1(L;\fso(V))$.

In summary we conclude with the metric version of Theorem
\ref{th:def-n-Leibniz}.

\begin{theorem}\label{th:def-metric-n-Leibniz}
  Infinitesimal metric deformations of a metric $n$-Leibniz algebra $V$ are classified
  by $HL^1(L;\fso(V))$ with $\fso(V)$ the representation of
  $L=V^{\otimes (n-1)}$ defined by \eqref{eq:EndVrep}.  The
  obstructions to integrating such an infinitesimal deformation live in
  $HL^2(L;\fso(V))$.
\end{theorem}

Finally, we may consider also deformations of a metric $n$-Lie
algebra $V$.  Now we have $L(V)= \Lambda^{n-1}V$ as explained in
Section \ref{sec:n-Lie} and we restrict to the subrepresentation
$\fso(V) \subset \End V$.  \emph{Mutatis mutandis} we arrive at the
following metric version of Theorem \ref{th:def-n-Lie}.

\begin{theorem}\label{th:def-metric-n-Lie}
  Infinitesimal metric deformations of a metric $n$-Lie algebra are
  classified by $H^1(C^\bullet)$, where $C^\bullet \subset
  CL^\bullet(L;\fso(V))$, for $L=\Lambda^{n-1}V$, is the subcomplex
  defined by
  \begin{equation}
    C^p = CL^p(L;\fso(V)) \quad \text{for $p\neq 1$, and}
  \end{equation}
  $C^1 \subsetneq CL^1(L;\fso(V))$ consists of those maps $\varphi: L
  \to \fso(V)$ whose associated $n$-linear map given by equation
  \eqref{eq:assoc-n-map} is totally skewsymmetric.  The obstructions
  to integrating an infinitesimal deformations live in
  $H^2(C^\bullet)$.
\end{theorem}

Similar remarks to those in Section \ref{sec:n-Lie} after Theorem
\ref{th:def-n-Lie} apply here as well.

\section{The case $n=3$}
\label{sec:n=3}

Due to their starring rôle in the construction of three-dimensional
superconformal Chern--Simons theories with matter, 3-Lie algebras and
more generally 3-Leibniz algebras deserve separate consideration.

\subsection{The Leibniz algebra in the Faulkner construction}
\label{sec:Faulkner}

As shown in \cite{Lie3Algs} all the metric 3-Leibniz algebras which
have appeared in the construction of three-dimensional superconformal
Chern--Simons theories with matter are special cases of a construction
due originally to Faulkner \cite{FaulknerIdeals}.  We will recall this
construction here and show how it too gives rise to a metric Leibniz
algebra.

Let $\fg$ be a real finite-dimensional Lie algebra with an
ad-invariant symmetric bilinear form $\left(-,-\right)$ and let $V$ be
a finite-dimensional faithful representation of $\fg$ with dual
representation $V^*$.  We will let $\left<-,-\right>$ denote the dual
pairing between $V$ and $V^*$.  Transposing the $\fg$-action defines
for all $v\in V$ and $\alpha \in V^*$ an element $\eD(v\otimes\alpha)\in
\fg$ by
\begin{equation}
  \label{eq:D-map}
  \left(X,\eD(v\otimes\alpha)\right) = \left<X \cdot v, \alpha\right>
  \quad\text{for all $X\in\fg$,}
\end{equation}
where the $\cdot$ indicates the $\fg$-action on $V$.  Extending $\eD$
linearly, defines a $\fg$-equivariant map $\eD : V \otimes V^* \to
\fg$, which as shown in \cite{Lie3Algs} is surjective because $V$ is a
faithful representation.  To lighten the notation we will write
$\eD(v,\alpha)$ for $\eD(v\otimes \alpha)$ in the sequel.  The
$\fg$-equivariance of $\eD$ is equivalent to
\begin{equation}
  \label{eq:FI-Faulkner}
  [\eD(v,\alpha),\eD(w,\beta)] =
  \eD(\eD(v,\alpha)\cdot w ,\beta) + \eD(w,
  \eD(v,\alpha)\cdot \beta)~,
\end{equation}
for all $v,w \in V$ and $\alpha ,\beta \in V^*$, where the dual action
$\eD(v,\alpha)\cdot \beta$ is defined by
\begin{equation}
  \label{eq:dualmodule}
  \left<w, \eD(v,\alpha)\cdot \beta\right> =
  -\left<\eD(v,\alpha)\cdot w, \beta\right>~.
\end{equation}
The map $\eD$ defines in turn a trilinear product
\begin{equation}
  \label{eq:3-product}
  \begin{aligned}[m]
    V \times V^* \times V &\to V\\
    (v,\alpha,w) &\mapsto \eD(v,\alpha) \cdot w~.
  \end{aligned}
\end{equation}

The fundamental identity \eqref{eq:FI-Faulkner} suggests defining a
bracket on $V \otimes V^*$ by
\begin{equation}
  \label{eq:LeibnizFaulkner}
  [v \otimes \alpha, w \otimes \beta] = \eD(v,\alpha)\cdot w \otimes
  \beta + w \otimes \eD(v,\alpha) \cdot \beta~,
\end{equation}
which would make $\eD$ into a morphism.  Indeed, we have the following

\begin{proposition}
  The bracket \eqref{eq:LeibnizFaulkner} turns $V \otimes V^*$ into a
  Leibniz algebra.
\end{proposition}

\begin{proof}
  We need only check the Leibniz identity \eqref{eq:LI}:
  \begin{equation}
    [u \otimes \alpha, [v\otimes\beta, w\otimes\gamma]] -
    [[u \otimes \alpha, v\otimes\beta], w\otimes\gamma] -
    [v\otimes\beta, [u \otimes \alpha, w\otimes\gamma]] \stackrel{?}{=}0~.
  \end{equation}
  We calculate each term in turn to obtain
  \begin{align*}
    [u \otimes \alpha, [v\otimes\beta, w\otimes\gamma]]
    &= [u \otimes \alpha, \eD(v,\beta)\cdot w \otimes \gamma + w \otimes
    \eD(v,\beta)\cdot \gamma]\\
    &= \eD(u,\alpha) \cdot \eD(v,\beta)\cdot w \otimes \gamma +
    \eD(v,\beta)\cdot w \otimes \eD(u,\alpha) \cdot\gamma\\
    & \qquad + \eD(u,\alpha) \cdot w \otimes \eD(v,\beta)\cdot \gamma + w
    \otimes \eD(u,\alpha) \cdot\eD(v,\beta)\cdot \gamma~,
  \end{align*}
  \begin{align*}
    [[u \otimes \alpha, v\otimes\beta], w\otimes\gamma]
    &= [\eD(u,\alpha)\cdot v\otimes \beta + v \otimes \eD(u,\alpha)\cdot
    \beta], w\otimes\gamma]\\
    &= \eD(\eD(u,\alpha)\cdot v,\beta) \cdot w \otimes \gamma + w \otimes
    \eD(\eD(u,\alpha)\cdot v,\beta) \cdot \gamma\\
    & \qquad + \eD(v,\eD(u,\alpha)\cdot \beta) \cdot w \otimes \gamma +
    w \otimes \eD(v,\eD(u,\alpha)\cdot \beta) \cdot \gamma~,
  \end{align*}
  and
  \begin{align*}
    [v\otimes\beta, [u \otimes \alpha, w\otimes\gamma]]
    &= [v\otimes\beta, \eD(u,\alpha)\cdot w \otimes \gamma + w \otimes
    \eD(v,\alpha)\cdot \gamma]\\
    &= \eD(v,\beta) \cdot \eD(u,\alpha)\cdot w \otimes \gamma +
    \eD(u,\alpha)\cdot w \otimes \eD(v,\beta) \cdot\gamma\\
    & \qquad + \eD(v,\beta) \cdot w \otimes \eD(u,\alpha)\cdot \gamma +
    w \otimes \eD(v,\beta) \cdot\eD(u,\alpha)\cdot \gamma~.
  \end{align*}
  Finally, putting it all together we find
  \begin{multline*}
    [u \otimes \alpha, [v\otimes\beta, w\otimes\gamma]] -
    [[u \otimes \alpha, v\otimes\beta], w\otimes\gamma] -
    [v\otimes\beta, [u \otimes \alpha, w\otimes\gamma]] \\
    = \left([\eD(u,\alpha), \eD(v,\beta)] - \eD(\eD(u,\alpha)\cdot
      v,\beta) - \eD(v,\eD(u,\alpha)\cdot \beta) \right)  \cdot w \otimes \gamma \\
    + w \otimes \left( [\eD(u,\alpha), \eD(v,\beta)] -
      \eD(\eD(u,\alpha)\cdot v,\beta) - \eD(v,\eD(u,\alpha)\cdot \beta)
    \right) \cdot \gamma~,
  \end{multline*}
  which vanishes by virtue of the fundamental identity
  \eqref{eq:FI-Faulkner}.
\end{proof}

Therefore the bracket \eqref{eq:LeibnizFaulkner} defines a (left)
Leibniz algebra structure on $V \otimes V^*$ making the map $\eD: V
\otimes V^* \to \fg$ into a Leibniz algebra morphism.  Notice that $V
\otimes V^* \cong \End V$ as vector spaces, but the induced Leibniz
algebra structure on $\End V$ is different in general from the Lie
algebra structure given by the commutator.

The vector space $V \otimes V^*$ has a natural $\fg$-invariant inner
product induced form the dual pairing between $V$ and $V^*$.  Under
the vector space isomorphism $V\otimes V^* \cong \End V$, this inner
product is simply the trace of the product of endomorphisms.  On
monomials, it is defined by
\begin{equation}
  \label{eq:FaulknerLIP}
  \left<v\otimes \alpha, w\otimes \beta\right> = \left<w,\alpha\right>
  \left<v,\beta\right>~,
\end{equation}
for all $v,w\in V$ and $\alpha, \beta \in V^*$, and on all of $V
\otimes V^*$ by extending linearly.  Since this inner product is
induced from the dual pairing, it is invariant under $\fg$, and hence
under the left Leibniz action of $V\otimes V^*$ on itself.

\begin{proposition}
  The Leibniz algebra $V\otimes V^*$ with bracket defined by
  \eqref{eq:LeibnizFaulkner} is metric with respect to the inner
  product defined by \eqref{eq:FaulknerLIP}.
\end{proposition}

\begin{proof}
  Let $X\in V\otimes V^*$ and let $\alpha,\beta \in V^*$ and $v,w \in
  V$. Then,
  \begin{align*}
    \left<[X, v\otimes\alpha], w\otimes \beta\right> &=
    \left<\eD(X) \cdot (v\otimes\alpha), w\otimes\beta\right>\\
    &= \left<\eD(X) \cdot v\otimes\alpha + v\otimes \eD(X) \cdot \alpha,
      w\otimes\beta\right>\\
    &= \left<\eD(X)\cdot v, \beta\right> \left<w,\alpha\right> + 
    \left<v, \beta\right> \left<w,\eD(X)\cdot \alpha\right>\\
    &= - \left< v, \eD(X)\cdot\beta\right> \left<w,\alpha\right> -
    \left<v, \beta\right> \left<\eD(X)\cdot w,\alpha\right>\\
    &= - \left<v\otimes\alpha, \eD(X) \cdot  w\otimes\beta + w \otimes
    \eD(X) \cdot \beta\right>\\
    &= - \left<v\otimes\alpha, \eD(X) \cdot  (w\otimes\beta)\right>\\
    &= - \left<v\otimes\alpha, [X,w \otimes \beta]\right>~.
  \end{align*}
\end{proof}

\subsection{3-Leibniz algebras arising from the real Faulkner construction}
\label{sec:cs3algebras}

A special case of the Faulkner construction recalled above is where
$V$ is a faithful unitary representation of $\fg$.  This means that
$V$ is a real, complex or quaternionic representation of $\fg$
possessing a $\fg$-invariant real symmetric, complex hermitian or
quaternionic hermitian inner product, respectively.  This gives rise,
respectively, to a real orthogonal, complex unitary or quaternionic
unitary representation of $\fg$.  As explained in \cite{SCCS3Algs}, we
may take the real case as fundamental and think of the complex and
quaternionic unitary cases as simply adding extra structure: an
invariant orthogonal complex structure for the complex case and an
anticommuting pair of such complex structures for the quaternionic
case.  As shown in \cite{Lie3Algs}, the real case corresponds
precisely to the metric 3-Leibniz algebras constructed by Cherkis and
Sämann in \cite{CherSaem}.  We briefly recall this construction here
in order to later set up the deformation theory of such algebras.

We will first briefly review the case of $(V,\left<-,-\right>)$ a real
inner product space admitting a faithful orthogonal action of a real
metric Lie algebra $\fg$.  The inner product on $V$ sets up an
isomorphism $\flat: V \to V^*$ of $\fg$-modules, defined by $v^\flat =
\left<v,-\right>$, with inverse $\sharp: V^* \to V$.  The map $\eD: V
\otimes V^* \to \fg$ defined by equation \eqref{eq:D-map} induces a
map $D: V \otimes V \to \fg$, by $D(v \otimes w) = \eD(v\otimes
w^\flat)$.  In other words, for all $v,w\in V$ and $X \in \fg$, we
have
\begin{equation}
  \label{eq:D-map-R}
  \left(D(v\otimes w),X\right) = \left<X \cdot v, w\right>~.
\end{equation}
It follows from the $\fg$-invariance of the inner product that
\begin{equation}
  \left(D(v \otimes w),X\right)  = \left<X \cdot v, w\right> = -
  \left<X \cdot w, v\right>  = - \left(D(w\otimes v),X\right)~,
\end{equation}
whence
\begin{equation}
  \label{eq:D-skew}
  D(v \otimes w) = - D(w\otimes v)~.
\end{equation}
This means that $D$ factors through a map also denoted $D: \Lambda^2V
\to \fg$.

Using $D$ we can define a 3-bracket on $V$ by
\begin{equation}
  \label{eq:3-bracket-R}
  [u,v,w] := D(u\wedge v) \cdot w~,
\end{equation}
for all $u,v,w \in V$.  The resulting 3-Leibniz algebra, which
appeared originally in \cite{FaulknerIdeals} but more recently in
\cite{CherSaem} in the context of superconformal Chern--Simons-matter
theories, satisfies the following axioms for all $x,y,z,v,w\in V$:
\begin{enumerate}
\renewcommand{\labelenumi}{(\alph{enumi})}
\item the \emph{orthogonality} condition
  \begin{equation}
    \label{eq:unitarity-R}
    \left<[x,y,z], w\right> =  - \left<z, [x,y,w]\right>~;
  \end{equation}
\item the \emph{symmetry} condition
  \begin{equation}
    \label{eq:symmetry-R}
    \left<[x,y,z], w\right> = \left<[z,w,x], y\right>~;
  \end{equation}
\item and the \emph{fundamental identity}
  \begin{equation}
    \label{eq:FI-R}
    [x,y,[v,w,z]] - [v,w,[x,y,z]] = [[x,y,v],w,z] + [v,[x,y,w],z]~.
  \end{equation}
\end{enumerate}
It follows from the orthogonality and symmetry conditions that
$[x,y,z] = - [y,x,z]$ for all $x,y,z \in W$, which is nothing but
equation \eqref{eq:D-skew}.  We will call such metric 3-Leibniz
algebras \textbf{Cherkis--Sämann} 3-algebras.  They have as a special
case the metric 3-Lie algebras appearing in the maximally
supersymmetric $N=8$ theory of Bagger--Lambert \cite{BL1,BL2} and
Gustavsson \cite{GustavssonAlgM2}, wherein the 3-bracket is totally
skewsymmetric.  Another special case of these 3-Leibniz algebras
corresponds to metric Lie triple systems, for which the 3-bracket
obeys $[x,y,z] + [y,z,x] + [z,x,y] =0$.  Metric Lie triple systems are
characterised by the fact that they embed into $\fg \oplus V$ as a
real metric $\ZZ_2$-graded Lie algebra and are in one-to-one
correspondence with pseudoriemannian symmetric spaces.

An easy consequence of the results in Section \ref{sec:Faulkner} is
that the Leibniz algebra $L(V) = \Lambda² V$ is metric relative to
the standard determinantal inner product:
\begin{equation}
  \label{eq:CSLIP}
  \left<u\wedge v, w \wedge z\right> =
  \left<u,w\right>\left<v,z\right> - \left<u,z\right>
  \left<v,w\right>~.
\end{equation}

\subsection{Deformation complex of Cherkis--Sämann 3-algebras}
\label{sec:deformations-CS}

Let us now set up the deformation theory of these metric 3-Leibniz
algebras, obtained via the Faulkner construction associated to a real
orthogonal representation of a metric Lie algebra.  As discussed
above, such an algebra consists of a real vector space $V$ and a
linear map $D:\Lambda^2V \to \fso(V)$ satisfying the fundamental
identity \eqref{eq:FI} and, in addition, the symmetry condition
\eqref{eq:symmetry-R}.  In the absence of the symmetry condition, the
deformation theory of those algebras are covered by the results in
Section~\ref{sec:metric-deformations} and in particular by
Theorem~\ref{th:def-metric-n-Leibniz}, but applied to the Leibniz
algebra $\Lambda^2 V$.  The symmetry condition requires special
consideration.  The situation here is analogous to that of metric
$n$-Lie algebras, except that instead of total skewsymmetry of the
bracket, the additional algebraic condition we are imposing is
equation \eqref{eq:symmetry-R}.  Following the discussion in
Section~\ref{sec:n-Lie}, we define a graded subspace $C^\bullet
\subset CL^\bullet(L;\fso(V))$, where $L=\Lambda^2V$, by $C^p =
CL^p(L;\fso(V))$ for $p\neq 1$ and $C^1 \subsetneq CL^1(L;\fso(V))$
consists of those $\varphi: \Lambda^2V \to \fso(V)$ such that
\begin{equation}
  \left<\varphi(u \wedge v) \cdot x, y\right> = 
  \left<\varphi(x \wedge y) \cdot u, v\right>~,
\end{equation}
for all $u,v,x,y \in V$.

\begin{lemma}\label{le:CS3-subcomplex}
  The subspace $C^\bullet$ so defined is a subcomplex of
  $CL^\bullet(L;\fso(V))$.
\end{lemma}

\begin{proof}
  We need only verify that the image of the differential $d:
  CL^0(L;\fso(V)) \to CL^1(L;\fso(V))$ actually lives inside $C^1$.
  To this end let $f \in \fso V = CL^0(L;\fso(V))$.  Let $X \in L$ and
  $\psi \in \fso(V) = CL^0(L;\fso(V))$ and consider $d\psi \in
  CL^1(L;\fso(V))$.  Then
  \begin{equation}
    d\psi(X) = -[\psi,X] = -[\psi,D(X)] + D(\psi \cdot X)~,
  \end{equation}
  whence
  \begin{align*}
    \left<d\psi(u \wedge v) \cdot x, y\right> &= -
    \left<[\psi,D(u\wedge v)] \cdot x, y\right> + \left< D(\psi \cdot
      (u\wedge v))\cdot x, y\right>\\
    &=  -\left<\psi \cdot D(u\wedge v) \cdot x, y\right> + \left<
      D(u\wedge v) \cdot \psi \cdot x, y\right>\\
    &\qquad + \left< D(\psi \cdot u \wedge v ) \cdot x, y\right> +
    \left< D(u\wedge \psi\cdot v)\cdot x, y\right>\\
    &= \left<D(u\wedge v) \cdot x, \psi \cdot y\right> + \left<
      D(u\wedge v) \cdot \psi \cdot x, y\right>\\
    &\qquad + \left< D(\psi \cdot u \wedge v ) \cdot x, y\right> +
    \left< D(u\wedge \psi\cdot v)\cdot x, y\right>\\
    &= \left<D(x\wedge \psi\cdot y) \cdot u, v\right> + \left<
      D(\psi \cdot x \wedge y) \cdot u,v\right>\\
    &\qquad + \left< D(x \wedge y ) \cdot \psi \cdot u, v \right> +
    \left< D(x\wedge y)\cdot u, \psi\cdot v\right>\\
    &= \left<D(\psi \cdot (x\wedge y)) \cdot u, v\right> + \left< [D(x
      \wedge y ) ,\psi] \cdot u, v \right>\\
    &= \left<D(\psi \cdot (x\wedge y)) \cdot u, v\right> - \left<
      [\psi, D(x \wedge y )] \cdot u, v \right>\\
    &= \left<d\psi(x \wedge y) \cdot u, v\right>~,
  \end{align*}
  whence $d\psi \in C^1$.
\end{proof}

In complete analogy to Theorem \ref{th:def-n-Lie}, we have the
following

\begin{theorem}\label{th:def-3-CS}
  Infinitesimal deformations of a metric $3$-Leibniz algebra $V$
  obtained by the real Faulkner construction are classified by
  $H^1(C^\bullet)$, where $C^\bullet \subset CL^\bullet(L;\End V)$ is
  the subcomplex defined above.  The obstructions to integrating an
  infinitesimal deformations live in $H^2(C^\bullet)$.
\end{theorem}

Similar remarks to those after Theorem \ref{th:def-n-Lie} apply here
as well.

\subsection{Some calculations}
\label{sec:calculations}

Since, in the absence of general theoretical results on the cohomology
of Leibniz algebras, calculations of deformations (or rigidity) of
metric 3-Leibniz algebras seem to be amenable only to explicit
solution of the cocycle and coboundary equations, one is inevitably,
albeit reluctantly, driven to work relative to a basis in such a way
that one can then harness the power of symbolic computation.

Let $V$ be an $N$-dimensional 3-Leibniz algebra with basis $(e_a)$.
Relative to this basis, the 3-bracket is give by the structure
constants $F_{abc}^d$ defined by
\begin{equation}
  [e_a,e_b,e_c] = F_{abc}^d e_d,
\end{equation}
where here and in the sequel we will employ the summation convention.

The corresponding Leibniz algebra $L = V \otimes V$ is
$N^2$-dimensional and has basis $e_{ab}:=e_a \otimes e_b$.  A
0-cochain $f \in CL^0(L;\End V) = \End V$ is given by a tensor $f_a^b$
defined by
\begin{equation}
  f(e_b) = f_a^b e_b~,
\end{equation}
whereas a 1-cochain $\varphi \in CL^1(L;\End V)$ is given by a tensor
$\varphi_{abc}^d$ defined by
\begin{equation}
  \varphi(e_{ab})(e_c) = \varphi_{abc}^d e_d~.
\end{equation}
Such a 1-cochain is 1-coboundary, $\varphi = df$, if and only if
\begin{equation}
  \label{eq:coboundarybasis}
  \varphi_{abc}^d = f_a^e F_{ebc}^d + f_b^e F_{aec}^d + f_c^e
  F_{abe}^d - F_{abc}^e f_e^d~,
\end{equation}
where it is a 1-cocycle, $d\varphi = 0$, if and only if
\begin{equation}
  \label{eq:cocyclebasis}
  \varphi_{cde}^g F_{abg}^f - F_{abe}^g \varphi_{cdg}^f + F_{cde}^g
  \varphi_{abg}^f - \varphi_{abe}^g F_{cdg}^f - \varphi_{abd}^g
  F_{cge}^f - \varphi_{abc}^g F_{gde}^f - F_{abc}^g \varphi_{gde}^f -
  F_{abd}^g \varphi_{cge}^f = 0~.
\end{equation}

Equations \eqref{eq:coboundarybasis} and \eqref{eq:cocyclebasis} are
not altered when we consider 3-Lie algebras, except that now
$\varphi_{abc}^d$ is totally skewsymmetric in $abc$.  As we remarked
in more generality after the proof of Lemma \ref{le:subcomplex}, we
see here explicitly that the coboundary $df$ is simply the action of
the endomorphism $f \in \fgl(V)$ on the tensor $F$, and hence if $F$
belongs to some submodule of $\fgl(V)$, so will $df$.

This has the following practical upshot for the computation of the
infinitesimal deformations.  To deform in a class of algebras larger
than the one the original algebra lies in, e.g., to deform a 3-Lie
algebra as a 3-Leibniz algebra, one simply relaxes the total
skewsymmetry of $\varphi_{abc}^d$ from the start.  The space of
coboundaries will not change, but the space of cocycles might be
enlarged, as one would expect.

To illustrate this, let us consider the unique simple euclidean 3-Lie
algebra, here denoted $S_4$.

\begin{example}[The Leibniz algebra of the simple euclidean 3-Lie
  algebra $S_4$]
  \label{eg:S4}
  Take $V=\RR^4$ with the standard inner product.  Let $(e_a)$, for
  $1\leq a \leq 4$, be an orthonormal basis.  The associated Leibniz
  algebra is $\Lambda^2 \RR^4$ with basis $(e_{ab}:=e_a\wedge e_b)$,
  for $1\leq a<b\leq 4$.  The bracket of the 3-Lie algebra is given by
  \begin{equation}
    [e_a,e_b,e_c] = \varepsilon_{abcd} e_d~,
  \end{equation}
  with the conventions that $\varepsilon_{1234} = +1$.  The bracket in
  the Leibniz algebra is given by
  \begin{equation}
    [e_{ab},e_{cd}] = \varepsilon_{abce} e_{ed} + \varepsilon_{abde} e_{ce}~.
  \end{equation}
  The Lie bracket in $\fg = \fso(4)$ is given by
  \begin{equation}
    [D(e_{ab}),D(e_{cd})] = \varepsilon_{abce} D(e_{ed}) +
    \varepsilon_{abde} D(e_{ce})~.
  \end{equation}
  Since $D$ has no kernel, it is an isomorphism of Leibniz algebras
  $\Lambda^2\RR^4 \to \fso(4)$.  Since $\fso(4)$ is Lie, so is
  $\Lambda^2\RR^4$.  The inner product in $V$ is such that the $e_a$
  are orthonormal, and this implies that in the Leibniz algebra
  \begin{equation}
    \left<e_{ab},e_{cd}\right> = \delta_{ac}\delta_{bd} -
    \delta_{ad}\delta_{bc}~,
  \end{equation}
  whereas in the Lie algebra of inner derivations
  \begin{equation}
    \left<D(e_{ab}),D(e_{cd})\right> = \varepsilon_{abcd}~.
  \end{equation}
\end{example}

An explicit calculation (made less painful using symbolic computation,
e.g., Mathematica) reveals that $S_4$ is rigid as a 3-Lie algebra,
whereas it admits a one-parameter deformation as a 3-Leibniz algebra
\begin{equation}
  [e_a,e_b,e_c]_t = \varepsilon_{abcd}e_d + t \left(\delta_{bc} e_a
    - \delta_{ac} e_b\right)~.
\end{equation}
It is interesting that this deformed 3-Leibniz algebra is already of
Faulkner type.  As such this deformation can be understood from the
Faulkner construction, as we now show.

For $t^2 \neq 1$, the Faulkner Lie algebra is $\fg \cong \fso(4) \cong
\fso(3) \oplus \fso(3)$ which admits, up to rescalings, a pencil of
$\fg$-invariant inner products on $\fg$.  The point $t=0$ corresponds
to the initial point in the deformation, namely the 3-Lie algebra
$S_4$ and corresponds as well to an inner product on $\fg$ which has
split signature.  For $t^2>1$ the signature of the inner product is
either positive-definite (for $t<-1$) or negative-definite ($t>1$).
As $t\to\pm\infty$ the algebra tends to the metric Lie triple system
associated to $S^4 = \SO(5)/\SO(4)$, thought of as a riemannian
symmetric space.  At the points $t=\pm 1$, the Faulkner Lie algebra is
isomorphic to $\fso(3)$: the selfdual $\fso(3)$ for $t=-1$ and the
antiselfdual for $t=+1$.  The inner product in either case is a
multiple of the Killing form: being positive-definite for $t=-1$ and
negative-definite for $t=+1$.

\subsection{Deformations of the Faulkner data}
\label{sec:deforming-faulkner}

The above example suggests that we ought to be able to understand
deformations of the metric 3-algebra in terms of deformations of its
Faulkner data.  Let us consider a metric 3-algebra of Faulkner type on
a finite-dimensional real vector space $V$ with symmetric inner
product $\left<-,-\right>$.  The Faulkner data is given by a metric
Lie algebra $\fg$ with ad-invariant inner product
$\kappa:=\left(-,-\right)$ and an embedding $\iota: \fg \to \fso(V)$.
Sylvester's Law of Inertia says that the signature of a nondegenerate
inner product cannot change under deformations, whence we may take the
inner product on $V$ and hence $\fso(V)$ to be rigid.  This means that
the structures getting deformed are the Lie bracket on $\fg$, the
inner product $\kappa$ on $\fg$ and the embedding $\iota$.  In the above
example we see that there are values of the deformation parameter
where $\fg$ drops dimension.  In order to take this into account, it
is convenient to fix the underlying vector space of $\fg$ but allow
$\kappa$ to be degenerate and $\iota$ to have nontrivial kernel.  Then the
true Faulkner Lie algebra is not $\fg$ but $\fg/\rad \kappa$, where
the radical
\begin{equation}
  \rad\kappa = \left\{X \in \fg\middle | \kappa(X,Y) =0~\forall Y \in
    \fg\right\}
\end{equation}
of $\kappa$ is an ideal of $\fg$, whence the quotient $\fg/\rad\kappa$
is a metric Lie algebra.   Indeed, it follows from equation
\eqref{eq:D-map-R} that $D(x\otimes y)$ is only defined modulo $\rad
\kappa$ and that if $X \in \rad\kappa$ then $X \in \ker \iota$.

By a \textbf{deformation of the Faulkner data} we mean a one-parameter
family consisting, for every $t$ in a neighbourhood of 0, of
\begin{itemize}
\item a linear map $[-,-]_t : \Lambda^2 \fg \to \fg$ subject
  to the Jacobi identity
  \begin{equation}
    \label{eq:Jac-F-t}
    [X,[Y,Z]_t]_t = [[X,Y]_t,Z]_t +  [Y,[X,Z]_t]_t~,
  \end{equation}
\item a bilinear form $\kappa_t: S^2\fg \to \RR$ subject to the
  ad-invariance condition
  \begin{equation}
    \label{eq:ad-inv-F-t}
    \kappa_t([X,Y]_t,Z) = - \kappa_t(Y,[X,Z]_t)~,
  \end{equation}
\item and a morphism $\iota_t: \fg \to \fso(V)$, whence subject to
  \begin{equation}
    \label{eq:hom-F-t}
    \iota_t[X,Y]_t = [\iota_tX,\iota_tY]~,
  \end{equation}
  where the bracket on the right-hand side is the one in $\fso(V)$.
\end{itemize}
Associated to this data there is a map $D_t:\Lambda^2 V \to \fg$
defined by
\begin{equation}
  \kappa_t(D_t(x \wedge y), X) = \left<\iota_t X \cdot x, y\right>~,
\end{equation}
for all $X \in \fg$ and $x,y\in V$, and a corresponding 3-bracket
\begin{equation}
  [x,y,z]_t = \iota_t D_t(x\wedge y) \cdot z~.
\end{equation}
Notice that $D_t(x\wedge y)$ is only defined up to $\rad\kappa_t$ and
that if $X \in \rad\kappa_t$, then $\iota_tX = 0$, whence the Faulkner Lie
algebra is $\fg_t/\rad\kappa_t$ and $\iota_t$ factors through an embedding
$\fg_t/\rad\kappa_t \to \fso(V)$.

The deformation equations \eqref{eq:Jac-F-t}, \eqref{eq:ad-inv-F-t}
and \eqref{eq:hom-F-t} are quadratic.  Linearising them around $t=0$
we obtain the linear equations which define an infinitesimal
deformation of the Faulkner data.  Let us write
\begin{equation}
  \begin{aligned}[m]
    [X,Y]_t &= [X,Y] + \sum_{k\geq 1} t^k \varphi_k(X,Y)\\
    \kappa_t(X,Y) &= \kappa(X,Y) + \sum_{k\geq 1} t^k \mu_k(X,Y)\\
    \iota_t(X) &= \iota(X) + \sum_{k\geq 1} t^k \lambda_k(X)~,
  \end{aligned}
\end{equation}
in terms of which, the infinitesimal deformations are given by
\begin{equation}
\label{eq:inf-def}
  \begin{aligned}[m]
    \varphi(X,[Y,Z]) + [X,\varphi(Y,Z)] - \varphi([X,Y],Z) - \varphi(Y,[X,Z]) - [\varphi(X,Y),Z] - [Y,\varphi(X,Z)] &= 0\\
    \kappa(\varphi(X,Y),Z) + \mu([X,Y],Z) + \kappa(Y,\varphi(X,Z))+\mu(Y, [X,Z]) &= 0\\
    \iota(\varphi(X,Y)) + \lambda([X,Y]) - [\iota(X),\lambda(Y)] - [\lambda(X), \iota(Y)] &= 0~,
  \end{aligned}
\end{equation}
where $\varphi:=\varphi_1$, $\mu:=\mu_1$ and $\lambda:=\lambda_1$.
The first equation is simply the cocycle condition for $\varphi \in
C^2(\fg;\fg)$.

A deformation is trivial if it is the result of the action of a
one-parameter subgroup of the general linear group $\GL(\fg)$.  For an
infinitesimal deformation this means that
\begin{equation}
  \label{eq:triv-def}
  \begin{aligned}[m]
    \varphi(X,Y) &=\psi([X,Y]) - [\psi(X),Y] - [X,\psi(Y)]\\
    \mu(X,Y) &= - \kappa(\psi(X),Y) - \kappa(X,\psi(Y))\\
    \lambda(X) &= - \iota(\psi(X))~,
  \end{aligned}
\end{equation}
for some $\psi \in \fgl(\fg)$.  The first equation simply says that
$\varphi = -d\psi$ for $\psi \in C^1(\fg;\fg)$.  One can easily check
that if $\varphi$, $\mu$ and $\lambda$ are given as in equations
\eqref{eq:triv-def}, then they also satisfy equations
\eqref{eq:inf-def}.

It is not clear, however, that computing deformations of the Faulkner
data is any easier than computing deformations of the 3-Leibniz
algebra itself; although if $\fg$ is semisimple, then one can do
better.  It may seem that this is a very special case, but notice that
if $V$ is positive-definite, then $\fg < \fso(V)$ is reductive, whence
the direct sum of a semisimple and an abelian Lie algebras.  So taking
$\fg$ semisimple is an important special case.

\begin{theorem}
  Let $(\fg,\kappa,\iota)$ be Faulkner data for a Cherkis--Sämann
  3-algebra $V$, where $\fg$ is semisimple.  Then only $\kappa$
  deforms and does so by rescaling the Killing form in each of its
  simple ideals.
\end{theorem}

\begin{proof}
  Indeed, if $\fg$ is semisimple, it is rigid and hence one can assume
  that $[-,-]_t$ is constant and equal to the original bracket.  The
  notion of trivial deformation now changes, of course, since
  $\GL(\fg)$ does not act on $\fg$ via automorphisms.  A trivial
  deformation is one which corresponds to the action of a
  one-parameter subgroup of $\Aut(\fg)$.  If $\fg$ is semisimple, this
  means a one-parameter subgroup of the adjoint group.

  Let $\fg = \fs_1 \oplus \dots \oplus \fs_k$ denote the decomposition
  of $\fg$ into its simple ideals.  Any ad-invariant symmetric
  bilinear form on $\fg$ takes the form
  \begin{equation*}
    \kappa = r_1 \kappa_1 + \dots + r_k \kappa_k~,
  \end{equation*}
  where $r_i \in \RR$ and $\kappa_i$ is the Killing form on $\fs_i$.
  Now the identity component of the adjoint group of $\fg$ (where
  one-parameter subgroups live) is the direct product of the identity
  components of the adjoint groups of each of the $\fs_i$.  These
  preserve the $\kappa_i$, whence trivial deformations
  actually leave $\kappa$ invariant.  The deformations of $\kappa$
  consist of changing the $r_i$, so that
  \begin{equation*}
    \kappa_t = r_1(t) \kappa_1 + \dots + r_k(t) \kappa_k~.
  \end{equation*}

  Finally we show that $\iota$ is also rigid.  Indeed as shown in
  \cite{MR0204575}, infinitesimal deformations of the morphism
  $\iota:\fg \to \fso(V)$ are classified by the Lie algebra cohomology
  space $H^1(\fg;\fso(V))$, where $\fso(V)$ becomes a $\fg$-module
  via $[\iota(X),-]$ for $X\in\fg$.  However for $\fg$ semisimple,
  $H^1(\fg,\fM) = 0$ for any $\fg$-module $\fM$ by the Whitehead
  Lemma.  Therefore $\iota$ is rigid.
\end{proof}

\section*{Acknowledgments}

The bulk of this paper was written while I was on sabbatical at the
University of Valencia in the second half of 2008.  I am grateful to
the University of Edinburgh for granting me the sabbatical absence and
to the University of Valencia for hospitality and support.  I would
like to thank José de Azcárraga for the invitation to Valencia and for
some initial conversations on the subject of this paper.  In fact this
paper is the result of my effort to identify the cohomology theory
resulting from a complex which he had written down and which later
emerged to be isomorphic to the one in \cite{MR2243339}.  I would also
like to thank Fernando Izaurieta for a very penetrating question,
which prompted me to think about and rediscover the relation between
3-Lie algebras and Leibniz algebras which permeate the results of this
paper.  It is also a pleasure to thank Sergey Cherkis and Neil Lambert
for their comments during a talk in Dublin about this work.  Last, but
certainly not least, I would like to thank Paul de Medeiros, Elena
Méndez-Escobar and Patricia Ritter for many useful and entertaining
3-algebraic discussions.

\bibliographystyle{utphys}
\bibliography{AdS,AdS3,ESYM,Sugra,Geometry,Algebra}

\end{document}